\documentclass[preprint]{JHEP}
\usepackage{amsmath,epsfig}
\usepackage{amssymb,amsfonts}
\usepackage{latexsym}
\renewcommand{\theequation}{\arabic{section}.\arabic{equation}}
\def\be{\begin{equation}}
\def\ee{\end{equation}}
\def\bea{\begin{eqnarray}}
\def\eea{\end{eqnarray}}
\def\RR{{\rm I\kern-1.6pt {\rm R}}}
\newcommand{\ha}{\frac{1}{2}}
\newcommand{\la}{\lambda}
\newcommand{\rc}{\nonumber\\}
\newcommand{\bear}{\begin{eqnarray}}
\newcommand{\eear}{\end{eqnarray}}

\def\spa{\;,\;}

\def\kkl{KKL$_4$\;}

\title{Non-critical holography and four-dimensional
\\  CFT's with fundamentals}
\author{
F. Bigazzi$^{a,b}$,
R. Casero$^c$,  A. L. Cotrone$^{d}$, E. Kiritsis$^{c,e}$, A. Paredes$^c$\\
~\\
$^a$ LPTHE, Universit{\'e}s Paris VI et VII, 4 place Jussieu \\
75252 Paris cedex 05, FRANCE\\
~\\
$^b$ INFN, Piazza dei Caprettari, 70, 
 I-00186  Roma, ITALY\\
~\\
$^c$ CPHT, Ecole Polytechnique, UMR du CNRS 7644, 
 91128 Palaiseau, FRANCE\\
~\\
$^d$ Departament ECM, Facultat de F{\'\i}sica, Universitat de Barcelona and \\Institut de
Fisica d'Altes Energies,
Diagonal 647, E-08028 Barcelona, SPAIN\\
~\\  
$^e$ Department of Physics, University of Crete, 
71003 Heraklion, GREECE\\
~\\
\texttt{\email{bigazzi@lpthe.jussieu.fr}, \email{casero@cpht.polytechnique.fr}, \email{cotrone@ecm.ub.es},  \email{paredes@cpht.polytechnique.fr}}}

\preprint{\hepth{0505140} \\ CPHT-RR 019.0305 \\ UB-ECM-PF-05/08\\ LPTHE-05-12}

\abstract{
We find non-critical string backgrounds in five and eight dimensions,
 holographically related to four-dimensional conformal field theories 
 with ${\cal N}=0$ and ${\cal N}=1$ supersymmetries. 
 In the five-dimensional case 
 we find an $AdS_5$ background metric for a string model 
 related to non-supersymmetric, conformal QCD with large number of 
 colors and flavors and discuss the conjectured existence of a conformal window
 from the point of view of our solution.
 In the eight-dimensional string theory,  we build a family of solutions 
  of the form $AdS_5\times \tilde{S}^3$ with $\tilde S^3$ a 
 squashed three-sphere. For a special value of the ratio $N_f/N_c$, the 
 background can be interpreted as the supersymmetric  near-horizon
limit of a system of color and flavor branes on $\mathbb{R}^{1,3}$ times
 a known four-dimensional generalization of the cigar. The $\mathcal{N}=1$ 
 dual theory with fundamental matter should have an IR fixed point only for 
 a fixed ratio $N_f/N_c$.  General features 
 of the string/gauge theory 
 correspondence for theories with fundamental flavors are also addressed.
}

\begin{document}


\section{Introduction}

The string/gauge theory correspondence \cite{malda,GKP,W} postulates the identity of  string theory on a non-trivial background and a related field theory living on the boundary of the space. In many important examples, most notably $\mathcal{N}=4$ super Yang-Mills and quiver superconformal gauge theories, the string theory dual lives in ten dimensions and there exists a range of parameters for which the background has small curvature compared to the string length, $\sqrt{\alpha'}$. Critical string theory is then reliably approximated by supergravity, faithfully reproducing the dual gauge theory at large 't Hooft coupling.

There are, however, important obstructions that forbid the construction of a weakly curved ten-dimensional string theory dual for every gauge theory. For many phenomenologically interesting field theories, critical string theory has too many degrees of freedom to be able to  reproduce the dynamics of the would-be dual field theory. In particular, the Kaluza-Klein states on some or all of the internal space directions couple to modes  that do not belong to the desired field theory. In order to decouple these unwanted modes, one has to take a limit in which the string theory geometry becomes highly curved and stringy dynamics is no longer negligible. This is the case for example for pure (super) Yang-Mills theories \cite{witten,KS, MN}, and it is expected for (large $N_c$) QCD.

This obstruction is not limited to confining theories.
In fact, there are also conformal theories for which a weakly coupled supergravity dual does not exist, as for example all conformal theories whose central charges $a$ and $c$ differ at leading order \cite{HS}. Among these, there are some of the most interesting conformal theories, such as non-supersymmetric conformal QCD, the IR fixed points of $\mathcal{N}=1$ theories containing fundamental matter, and $\mathcal{N}=2$ superconformal QCD.

For all these theories it is expected that the string dual is a non-critical string theory.
 Of course it is in general not known how to handle string theory off criticality. Moreover, in principle, the gravity approximation to non-critical string theory is not reliable because of the unavoidable large curvature of its background. Nevertheless, the gravity approximation has proven, from the first days of the string/gauge theory correspondence, to give remarkable qualitative insights, in particular for highly symmetric examples \cite{cave}. More specifically, string theory on $AdS$ vacua dual to conformal theories is believed to be subject to stringy corrections which do not spoil the solution but only affect its parameters \cite{Poly}. This is the reason why it is very fruitful to study examples of non-critical strings/conformal gauge theory pairs: the quest for a string theory dual of QCD or pure Yang-Mills necessarily requires the comprehension of all mechanisms on which the non-critical string/gauge correspondence is based, 
 which can be better understood starting from the study of  simpler $AdS$/CFT pairs.

Some examples of non-critical $AdS$/CFT pairs for theories with flavors have recently been studied \cite{KlebMald,Ali}\footnote{For other recent non-critical solutions see also \cite{KupSon, KupSon2}.}. In particular \cite{KlebMald} considers $\mathcal{N}=1$ QCD in Seiberg's conformal window. This example is particularly interesting because the IR field theory is conformal only when the ratio of the number of fundamental flavors to the number of colors is in the range $3/2<N_f/N_c<3$. Therefore fundamental flavors essentially contribute\footnote{Understanding flavors in string
duals beyond the quenched
approximation introduced in \cite{KarKatz} is still an open problem in ten dimensions, see \cite{backr} for works in this direction.} to the string background being $AdS$.
The setting considered in \cite{KlebMald} is  six-dimensional string theory on $\mathbb{R}^{1,3}\times SL(2,\mathbb{R})/U(1)$; the color Yang-Mills theory lives on D3-branes placed at the tip of the cigar (the $SL(2,\mathbb{R})/U(1)$ coset manifold), while flavors are introduced by means of space-time filling uncharged D5-branes. The effect of the flavor branes is encoded by the addition of  a Dirac action to the closed string effective action.
This setting has been further studied in \cite{FotoNiarPrez,Murthy}.

In this paper we further explore some new directions in the context of the
non-critical  gauge/gravity duality. Backgrounds with a non-trivial dilaton play
a crucial role in this setting, because they provide good vacua for
lower-dimensional string theories. D-branes placed on these backgrounds give
rise to gauge theories in a way similar to what happens in the
context of critical theories. We  address two different examples of 
non-critical
holographic pairs, involving in one case a five-dimensional string theory, and
in the other an eight-dimensional one. 
As usual for non-critical solutions, the backgrounds we build have 
 curvature of order one in units of $\alpha'$. We focus on duals
 of conformal field theories, so that,
 as explained above, 
 the Anti de Sitter structure of our solutions ensures that the
 two-derivative approximation  captures the qualitative features of 
 the full string models.

The five-dimensional backgrounds we find are a class of $AdS_5$ solutions\footnote{For the connection 
of higher spin theories in $AdS_5$ with gauged vector models, see \cite{Schnitzer}.}.
They can be thought of as
originating from the back-reaction of color D3 and 
flavor, space-time-filling, uncharged D4-branes on a linear 
dilaton background. The solutions have a natural interpretation as 
the duals of non-supersymmetric QCD with a large number of colors 
and massless flavors, in the conjectured conformal window 
\cite{BanksZaks}. 
We also analyze the spectrum of mesons by studying the fluctuations of 
the world-volume tachyon and gauge field of the flavor D4-branes.
Since the conformal dimension associated
 to these operators depends on $\frac{N_f}{N_c}$,
we argue that the limits of the conformal window should appear
when some of these modes become tachyonic or reach the unitarity bound.
However, a quantitative check is not possible due to 
${\cal O}(1)$  string corrections.

For the 
eight-dimensional case, we consider a vacuum which is the direct 
product of four-dimensional flat Minkowski and a four-dimensional 
K{\"a}hlerian space constructed in~\cite{KKL}, denoted by \kkl in 
the following.
This space has $U(2)$ isometry and it  originates 
from an ${\cal N}= 2$ world-sheet SCFT. The number of space-time 
supercharges for the string model on the above vacuum is eight.
This vacuum is a natural higher dimensional generalization of the 
two-dimensional cigar. 

We look for an eight-dimensional background which can be 
interpreted as being sourced by $N_c$ color branes with 
D3-charge  placed  at the tip of \kkl and $N_f$ suitably 
distributed, uncharged flavor D5-branes. The back-reacted geometry
turns out to be of the form $AdS_5\times~\!\! \tilde S ^3$, where
the internal $\tilde S ^3$ is a squashed three-sphere, with 
an $SU(2)\times U(1)$ isometry.
Although we build an infinite set of 
such solutions (depending on $N_f$ and $N_c$), we find 
that there is a special  value
of $\frac{N_f}{N_c}$, and present arguments that suggest
that this is the only supersymmetric solution of the set. 
The exact form of the ${\cal N}=1$ SCFT dual to the above particular 
solution is still to be found.

Notice that non-critical string theory in eight dimensions 
is particularly
interesting since, as  argued in section \ref{sect: finter}, 
we expect the
duals of ${\cal N}=2$ SYM and ${\cal N}=2$ SQCD to 
be constructed by placing branes on  $\mathbb{R}^{1,5}\times cigar$.
However, the fact that the full $SU(2)\times U(1)$ R-symmetry of such
gauge theories may not be reflected in the geometry hinders the computation
of the back-reacted solution. We leave this problem for future work.

On the same line, another eight-dimensional vacuum
 that would deserve attention is $\mathbb{R}^{1,1} \times 
 \mathrm{KKL}_6$.
 One can think of placing color and flavor branes in it in order
 to build the dual of an ${\cal N}=(1,1)$ gauge theory with
 fundamentals in 1+1 dimensions. 
 The back-reacted solution should be of the form $AdS_3\times X_5$
 where $X_5\sim SU(3)\times U(1)/(SU(2)\times U(1))$ 
 is a symmetric space with U(3) isometries.
 Its study, however, is beyond the
 scope of the present paper.

The plan of the paper is as follows.
In section \ref{CQCD} we deal with the five-dimensional setup. 
In section \ref{set+sol} 
we discuss the $AdS_5 \times \tilde S^3$ solutions.
We also study (part of) the spectrum of fluctuations 
with respect to small perturbations of the metric and dilaton and we 
discuss their stability.
In section \ref{sect BPS} we give some evidence signaling the
special $\frac{N_f}{N_c}$ ratio. In particular,
we employ a first order formalism that was 
first developed in the study of domain wall solutions in five-dimensional
 supergravity \cite{Freedman:1999gp, nonsusy}
and show that only for a particular
 value of this ratio it is possible to write an explicit,
  simple superpotential. 
We also build a flow from this $AdS_5\times \tilde S ^3$ 
solution to the string vacuum described above. 
Section \ref{sect: finter} is devoted to
 discussing  the open problems in the field theory interpretation of the above particular
 solution.

In section \ref{gen probls} we consider some general problems of 
the non-critical $AdS$/CFT correspondence, of which this paper 
gives some examples.
In particular, by examining field theory central charges, we stress
 the fact that generic superconformal theories with fundamental 
 matter and $N_f$ of the same order as $N_c$ cannot have a weakly 
 coupled supergravity dual. The same considerations should extend 
 to non-supersymmetric theories as well.

We conclude with a summary and outlook in section 
\ref{summary}. In four appendices we collect some results 
and checks on the solutions built in the main text.
In particular, in appendix~\ref{appendi} we discuss the 
semi-classical embedding in the \kkl vacuum of the branes we use to build 
the $AdS_5\times \tilde S ^3$ solutions.

\section{The five-dimensional solutions and conformal QCD}\label{CQCD}
In this section we will present the simplest conformal solution incorporating the
back-reaction of flavor branes.
It is a simple generalization of the $AdS_5$ solution found by Polyakov
 in \cite{cave}, but the inclusion of flavors provides some  interesting new physics.
The dual field theory is expected to  be simply QCD with enough flavors in the 
fundamental (and anti-fundamental) to be at a conformal point 
\cite{BanksZaks}.

The string-frame action in 5d is \cite{ferretti2,Poly} (with $\alpha'=1$):
\bea\label{ac}
S = \frac{1}{2\kappa^2_{(5)}} \int d^5x \sqrt{-g_{(5)}}\Big( e^{-2\phi}\big(R + 4 (\partial_\mu \phi)^2 + \mathfrak{c}\big)  - F_{(5)}^2 - 2Q_f e^{-\phi}\Big)\ ,
\eea
where $\kappa_{(5)}$ is the five-dimensional Newton constant, $\mathfrak{c}=10-D=5$ is the non-criticality constant 
needed to cancel the central charge and
preserve the world-sheet conformal symmetry, $F_{(5)}^2\equiv \frac{1}{5!}F_{\mu\nu\rho\sigma\lambda}F^{\mu\nu\rho\sigma\lambda}\sim Q_c^2$ with $Q_c$ being proportional to the rank $N_c$ of the gauge group, and $Q_f$ is proportional to the number of fundamental flavors $N_f$. We are thinking about three-branes  and space-time filling, uncharged four-branes giving the \mbox{$Q_f$-dependent} term.
The latter are the analogue of the uncharged 
five-branes introduced in \cite{KlebMald}.
In that case, they have been argued in \cite{FotoNiarPrez} to be ordinary 
branes double-sheeted on the cigar.
Here, on the other hand,  the four-branes should be most probably 
viewed as brane-antibrane pairs in order to be uncharged.

We will search for a conformal solution having constant dilaton and string frame metric: 
\be
ds^2= e^{2kr}dx_{1,3}^2 + dr^2\ .
\ee
The equations of motion for the constant dilaton ansatz are:
\bea
&& R= -\mathfrak{c} + Q_f e^{\phi}\ ,\label{erre}\\
&&R_{\mu\nu} = \frac{g_{\mu\nu}}{2}( R - 2Q_f e^{\phi}+ \mathfrak{c} ) + e^{2\phi} \frac{1}{4!}(F_{\mu \mu_2 ... \mu_5}F^{\ \mu_2 ... \mu_5}_{\nu}-\frac{g_{\mu\nu}}{10}F_{\mu_1 ... \mu_5}F^{\mu_1 ... \mu_5})\ ,\label{ricci1}\\
&&\partial_{\mu}\Big( \sqrt{-g_{(5)}} F^{\mu \mu_2 ... \mu_5} \Big)=0\ .\label{effe}
\eea
These equations have a simple $AdS_5$ solution with five-form 
$F_{(5)} = Q_c Vol(AdS_5)$.
In fact, the dilaton and Ricci tensor read:
\be
e^{\phi}=\frac{\sqrt{49Q_f^2+40 \mathfrak{c} Q_c^2}-7Q_f}{10Q_c^2}\label{dilaton}\ , \qquad\qquad
R_{\mu\nu} = -\frac{g_{\mu\nu}}{2} \left(Q_f e^{\phi} + Q_c^2 e^{2\phi}\right)\ ,
\ee
that is, the space is of constant curvature with radius $R^2_{AdS}=1/k^2$: 
\be\label{radQCD}
R^2_{AdS}=\frac{200 Q_c^2}{10 \mathfrak{c} Q_c^2+7 Q_f^2-Q_f\sqrt{40 \mathfrak{c} Q_c^2+49 Q_f^2}}\ .
\ee
The $R^2_{AdS}$ varies monotonically between $20/\mathfrak{c}$ and $28/\mathfrak{c}$, when $Q_f/Q_c$ varies from zero to infinity.
For large $Q_c$ and $Q_f$ the dilaton is always small, while the 't Hooft coupling $\lambda \sim e^{\phi}Q_c$ is of order one.
For $Q_f=0$ we end up with  the Polyakov solution \cite{cave}.

The interesting feature of this solution is that, unlike the solution in \cite{KlebMald}, the radius is not independent of $Q_c$ and $Q_f$.
This fact implies that the mass of the open string tachyon of the system of D4-anti D4 branes depends on $Q_c$ and $Q_f$ too, so that it can give constraints on their allowed values, as we will discuss in section \ref{fluct}.

The spectrum of five-dimensional string theory includes an 
axion and a one-form potential \cite{ferretti2}.
In fact, a constant axion can always be included 
in the solution above. It is as usual dual to the $\theta_{YM}$ term.
The baryonic $U(1)_B$ should be given by the one-form potential; 
the latter couples to a D0-brane which should be dual to  the baryon vertex, but the details of this identification require a deeper analysis.
The axial $U(1)_A$ can be identified with the symmetry of the complex tachyon on the D4 anti-D4 pairs.
The non-abelian $SU(N_f)\times SU(N_f)$ flavor symmetry is as usual the gauge symmetry on the D4's.

\subsection*{Inclusion of the closed string tachyon}\label{addtac}

Generically, the spectrum of non-critical closed strings contains (at least) 
one tachyonic scalar~$T$.
Since $T$ couples to the other fields, it should be included in the gravity action.
The action is a simple generalization of (\ref{ac}):
\be\label{ac0}
S  = \frac{1}{2\kappa^2_{(5)}} \int d^5x \sqrt{-g_{(5)}}\Big( e^{-2\phi}\big( R + 4 (\partial_\mu \phi)^2 + \mathfrak{c}(T) - \frac12 (\partial_\mu T)^2\big) - f(T)F_{(5)}^2 - 2Q_f g(T)e^{-\phi}\Big)
\ee
where $\mathfrak{c}(T)=5+\frac38 T^2 + ...$, $f(T)= 1 + O(T)$ and $g(T)= 1 + O(T)$ are known only up to the first few orders in their expansion around the zero-tachyon value.
There are linear terms in the tachyon in $f(T)$ and $g(T)$ for Type 0 theories  but not for Type II theories. 
As usual, we have to assume that the tachyon expectation value is small to keep it under control.

The equations of motion (\ref{erre})-(\ref{effe}) have the obvious generalization given by the inclusion of the tachyon factors, but still have the
 $AdS_5$ solution, with constant tachyon $T=T_0$ given by its equation of motion (here the prime means derivative w.r.t. the tachyon field):
\bea\label{tachion}
0 = \mathfrak{c}'(T_0)e^{-2\phi} - 2g'(T_0)Q_fe^{-\phi} -f'(T_0)F_{(5)}^2\ .
\eea
Type II theories admit the $T_0=0$ solution above.
In Type 0 theories this is not the case, but given a 
solution of (\ref{tachion}), one has to simply replace 
in the solution (\ref{dilaton}), (\ref{radQCD}), $\mathfrak{c}=5$ 
with $\mathfrak{c}(T_0)$, $Q_c^2$ with $Q_c^2 f(T_0)$ and $Q_f$ with $Q_fg(T_0)$.

\subsection{Fluctuations of the flavor branes}\label{fluct}

We will now briefly consider the quadratic fluctuations of the flavor branes, which will give us information on the spectrum of mesons in the dual field theory.
The fluctuations will be limited to the gauge fields on the branes and to the open string tachyon of the brane-anti brane pairs.
We will concentrate on the Abelian case, that is, a single D4-anti D4 brane pair.
The action for small open string tachyon $\tau$ is \cite{sen}:
\begin{equation}\label{azionetachy}
S=-\int d^{5}\xi\, V(|\tau|)e^{-\phi}(\sqrt{-det X^1}+\sqrt{-det X^2})\ ,
\end{equation}
with:
\be
V(|\tau|)=1-\frac{\pi}{2} |\tau|^2 + ...\ , \qquad   X_{ab}^{1,2}=g_{ab}^{1,2}+2\pi F_{ab}^{1,2} + \pi (\partial_a \tau (\partial_b \tau)^* + \partial_b \tau (\partial_a \tau)^*)\ .
\ee
Since the tachyon enters already quadratically, its fluctuations do not mix with the ones of the gauge fields.
Moreover, the two gauge fields $A^{1,2}$ are also 
decoupled to this order and have of course the same fluctuations.
Therefore, the fluctuating fields are just two free massless gauge fields 
(the $A_r$ component does not give a regular mode dual to a meson) and one massive 
(with bare mass $m^2=-1/2$) free complex scalar.
The conformal dimensions of the dual operators are given by  the standard AdS/CFT formula: 
$\Delta = 3$ for the gauge fields and
\be\label{dimtac}
\Delta_\pm = 2 \pm \sqrt{4-\frac{R^2_{AdS}}{2}}
\ee 
for the complex tachyon.
Note that the latter is no more tachyonic, due to the curvature of the background, if $R^2_{AdS}\leq 8$.
The boundary value $R^2_{AdS}= 8$ corresponds to the dimension $\Delta=2$.
As we decrease  $R^2_{AdS}$, we can have the two possible conformal dimensions $\Delta_\pm$ until we reach the value $R^2_{AdS}=6$, where $\Delta_-$ hits the unitarity bound (and $\Delta_+=3$).
For smaller values of the radius we have only $\Delta_+>3$.

The interpretation of this spectrum is straightforward.
The gauge fields are dual to the vector mesons.
The vectorial current is exactly conserved, so the dimension is protected by conformal symmetry and it is equal to the free one, namely $\Delta=3$.

For what concerns the fluctuations of the complex tachyon, they are dual to the scalar mesons.
The dimensions of these fields are not protected and differ from the free ones.
Conformal QCD is expected to have a conformal window as its ${\cal N}=1$ counterpart.
The supposed conformal window for this theory starts at a ratio $N_f/N_c$ which lies between $5/3\sim 1.7$ and $4$, and it ends surely before (or at) $N_f/N_c=11/2=5.5$ (see for example \cite{strassler}).
In analogy with the ${\cal N}=1$ theory, the dimensions of the scalar mesons are expected to violate the unitarity bound when we try to go outside this conformal window.

In fact, and unlike the solution dual to ${\cal N}=1$ SQCD \cite{KlebMald}, we find that these dimensions depend on $N_f$ and $N_c$ through $R^2_{AdS}$ (cfr. (\ref{radQCD})).
In this sense, this solution seems to give a confirmation of the expectations above.
Unfortunately, the fact that the gravity solution gets corrections of order one to the radius of $AdS$ (as usual for non-critical solutions) does not allow us to perform a meaningful numerical check of this picture.
In particular, from (\ref{radQCD}) we get $4 \leq R^2_{AdS}\leq 28/5 = 5.6$ for all values of $Q_f/Q_c$, so that the only conformal dimension is $\Delta_+$ (\ref{dimtac}) and it is always greater than three, thus not giving any conformal window. 
The addition of the closed string tachyon, if we insist in keeping it small, cannot improve substantially the picture, since it only gives a small correction to the numbers in $R^2_{AdS}$ (\ref{radQCD}).
We nevertheless expect that the complete (string corrected) solution 
will single out 
the values of $N_f/N_c$ corresponding to the conformal window \cite{KlebMald}.
We can in fact expect various types of corrections to the numbers we get.
The first comes from a non-small closed string tachyon value, shifting $\mathfrak{c}(T_0)$, $f(T_0)$, $g(T_0)$ and then $R^2_{AdS}$ by a relevant amount.
Moreover, the latter can well be changed by $\alpha'$ corrections.
Finally, we can expect that the effective mass of the open tachyon will get additional $N_c$, $N_f$ dependence from some higher order couplings we neglected, since this is the way it is expected to single out the conformal window in the ${\cal N}=1$ case \cite{KlebMald}.

\section{$AdS_5 \times \tilde S^3$: setup and solutions}\label{set+sol}

In this section we look for the solution obtained by considering
the back-reaction of color and flavor branes in 
an eight-dimensional  supersymmetric non-critical background, which can also be regarded as the simplest
generalization of the cigar.

\subsection{The setup}\label{setup}
In \cite{KKL} a sequence of solutions, KKL$_{2n}$, to the $2n$ dimensional string 
equations was found. They are K{\"a}hlerian spaces of complex dimension $n$
and isometry $U(n)$. They have a non-trivial dilaton, but the 
string coupling is bounded  from above. The first space in the series, 
KKL$_2$, coincides with  the 
well-known $SL(2,\mathbb{R})/U(1)$ coset CFT background, i.e. the 
2d cigar with $U(1)$ isometry. 
By suitably placing D-branes on 
$\mathbb{R}^{1,3}\times SL(2,\mathbb{R})/U(1)$, 
one can study the duals of ${\cal N}=1$ SQCD 
\cite{KlebMald,FotoNiarPrez} and ${\cal N}=1$ SYM \cite{Murthy}.

Here we consider the 
eight-dimensional vacuum $\mathbb{R}^{1,3}\times\, $\kkl:
\be
ds_8^2=dx_{1,3}^2 + ds_{\rm KKL_4}^2\,\,,
\label{8dvacuum}
\ee
where:
\bear\label{n2metric}
ds_{\rm KKL_4}^2&=&\frac{1}{ 4f_2(\zeta)} d\zeta^2 +\frac{\zeta}{4} (d\theta^2 + \sin^2\theta\, d\varphi^2)
+\frac {f_2(\zeta)}{ 4}(d\psi+\cos\theta \, d\varphi)^2\,\,,\rc
\phi&=&-\frac{\zeta }{ K} + {\rm const} \,\,.
\eear
The ranges for the angles are
$\theta \in [0,\pi]$, $\varphi \in [0,2\pi)$, $\psi \in [0,4\pi)$
and the function $f_2(\zeta)$ reads:
\be\label{funct f2}
f_2(\zeta)=\frac{K^2}{2}\,\frac{ 
e^{-\frac{2\zeta}{K}}-1 + \frac{2\zeta}{K}}{\zeta}\,\,.
\ee
For any value of $\zeta$,
the angular part of the \kkl metric 
describes a squashed 
three-sphere, also called Berger sphere. 
This homogeneous and anisotropic
space, which we will denote by $\tilde S^3$, 
can be described as a coset manifold $\frac{SU(2)\times U(1)}{U(1)}$ 
or as an $S^1$-fibration over $S^2$. It 
has the same topology as the round three-sphere but
its isometry group is just $SU(2)\times U(1)$. 
A review of several features of its geometry can be
found in \cite{Zoubos}.

In (\ref{funct f2}) we fixed an integration constant so that the metric is
smooth at the tip $\zeta =0$~\cite{HoriKap}. In fact, near that point, the
metric  reduces to flat $\RR^{4}$.
The constant $K$ is related to the background central charge that
must be canceled.
As the central charge deficit of the \kkl
space is:
\be
\delta c=\frac{24}{K}\,\,,
\ee
if one takes the product $\mathbb{R}^{1,D-5}\times\,$\kkl, it is necessary to
set:
\be\label{cc}
K=\frac{16}{10 -D}\,\,,
\ee
and therefore $K=8$ in the case of our interest. 

The interpretation of the \kkl background as the target space metric 
of a 2d ${\cal N}=~\!\!2$ CFT is discussed in \cite{HoriKap}. In that paper 
it is also shown that  the \kkl model with constant $K=4m$ emerges in 10d from
 a set of $m$ NS5 branes wrapped on the two-cycle of a $CY_2$. This 
 preserves  eight space-time supercharges.  
Our vacuum should thus preserve the same amount of supercharges.

To introduce colors, we place D3 branes at the tip of (\ref{n2metric}). This should break supersymmetry by 
one half, leaving ${\cal N}=1$ supersymmetry in 4d 
(which, as usual, can be enhanced to $\mathcal{N}=1$ superconformal 
symmetry if the solution contains an $AdS_5$ factor). To introduce flavors,
we add a stack of $N_f$ D5 branes filling all the non-compact directions
of (\ref{8dvacuum}) and wrapping a contractible one-cycle inside the 
 $\tilde S^3$.
Concretely, we take the flavor branes
 to be extended along $x_0, \dots, x_3, \psi$ and $\tau$, where we 
 have introduced a new  radial coordinate
 $\tau\in (-\infty,\infty)$ ($\tau\to -\infty$
corresponds to the tip of \kkl and is the point where the D3s are sitting).
These D5-branes are extended up to $\tau\to\infty$, so the 
gauge fields coming from the D5-D5 open
strings are decoupled from the theory living on the  D3-branes.
The D3-D5 open string degrees of freedom are quarks transforming
in the fundamental representation of the gauge group. 
In order to have gapless
D3-D5 strings and therefore massless quarks,  we
take the D5 to be extended also up to $\tau\to -\infty$.
Moreover, in order
to preserve the $SU(2)$ isometry of (\ref{n2metric}),
 they are homogeneously smeared on the transverse $S^2$ parameterized
 by $\theta$ and $\varphi$.
 Notice that this is only possible in the
limit $N_f\to \infty$. 
The brane configuration is summarized in
table~\ref{branescheme}.
\begin{table}[ht]
\begin{center}
\begin{tabular}{|c|c|c|c|c|c|c|c|c|}
\multicolumn{1}{c}{ }
&
\multicolumn{5}{c}{ }
&
\multicolumn{3}{c}{$\overbrace{\phantom{\qquad\ \ 
\ \ \,}}^{\tilde S^3}$}\\
\hline
\multicolumn{1}{|c|}{ }
&
\multicolumn{4}{|c|}{$x_{1,3}$}
&\multicolumn{1}{|c|}{$\tau$}
&\multicolumn{1}{|c|}{$\psi$}
&\multicolumn{1}{|c|}{$\theta$}
&\multicolumn{1}{|c|}{$\varphi$}\\
\hline
D3 &$-$&$-$&$-$&$-$&$\cdot$&$\cdot$&$\cdot$&$\cdot$\\
\hline
D5 &$-$&$-$&$-$&$-$&$-$&$\bigcirc$&$\sim $&$\sim$\\
\hline
\multicolumn{1}{c}{ }
&\multicolumn{4}{c}{$\underbrace{\qquad\qquad
\qquad}_{\textrm{Gauge theory}}$}
&\multicolumn{4}{c}{ }
\end{tabular}
\caption{Brane configuration in the general setup.
A line $-$ means that the brane spans a non-compact dimension, a point
$\cdot$ that it is point-like in that direction, a circle
$\bigcirc$ that it wraps a compact cycle and a 
tilde $\sim$ that, although
branes are not extended in that direction, they are uniformly distributed
there.
\label{branescheme}}
\end{center}
\end{table}

The action we consider is the (bosonic) two-derivative approximation to the 
low energy eight-dimensional non-critical IIA superstring theory\footnote{
We denote by IIA theory, in any dimension, the string theory whose 
RR sector is the tensor product of two spinors of opposite chirality.
In eight dimensions, this gives field-strengths that are one-, three-, five- and seven-forms.}. 
Keeping just the $F_{(5)}$ out of all possible RR field strengths,
we write the following string frame action:
\be\label{d=8 action}
S=\frac{1}{2 \kappa_{(8)}^2}\int d^8x \sqrt{-g_{(8)}}
\left( e^{-2\phi}
\left(R+4(\partial_\mu \phi)^2 + 2\right) - 
F_{(5)}^2 \right) + S_{flavor}\,\,.
\ee

We  look for eight-dimensional solutions with the same $SU(2)\times U(1)$ 
isometry as \kkl, which can be interpreted as the geometry produced by 
placing on  (\ref{8dvacuum}) the configuration of branes described above.
The natural ansatz for the back-reacted metric is (in string frame):
\be
ds_8^2=e^{2\lambda}\,dx_{1,3}^2+d\tau^2+
e^{2\nu}(d\theta^2 + \sin^2\theta\, d\varphi^2)+e^{2f}
(d\psi+\cos\theta\, d\varphi)^2\spa
\label{8dmetric}
\ee
where, in principle, the functions $\la$, $\nu$, $f$ depend on $\tau$.

The stack of $N_c$ D3 branes that supports
the gauge theory is extended in
the $x_i$-directions and is a source for 
the five-form RR field strength. The following ansatz for $F_{(5)}$
guarantees that  its equation of motion is satisfied:
\be
F_{(5)}=Q_c \,
e^{4\la -2\nu -f} \,dx^0\wedge\ldots\wedge dx^3\wedge d\tau\,\,,
\label{RRF5}
\ee
where the constant $Q_c$ is $N_c$ up to some normalization factor.

The term $S_{flavor}$ in (\ref{d=8 action}) is the action
for the flavor branes.
We will consider their back-reaction on
the geometry in a fashion similar to \cite{KlebMald}.
 Their  contribution 
is given by the Dirac action:
\be
S_{flavor}=-T_5 \sum_{i=1}^{N_f} \int_{{\cal M}_6}
 d^6\xi \,e^{-\phi}\,\sqrt{-\hat g_{(6)}} \spa
 \label{sflavor1}
\ee
where ${\cal M}_6$ is the brane world-volume and $\hat g_{(6)}$
stands for the determinant of the pull-back of the metric on
such world-volume.

Because of the smearing, we can substitute
the sum in (\ref{sflavor1})
by an integral over the space transverse to the D5 branes: defining
a constant density along the transverse $S^2$, we have
$\sum_{i=1}^{N_f} \to {N_f \over 4\pi}
\int d\theta d\varphi \sin\theta$.
We arrive at the expression\footnote{Notice that the difference from taking D7 flavor branes which would
fill the whole eight-dimensional
space-time is the $e^{2\nu}$ in the denominator. No
solutions of the type we are searching for can be found in this case;
we will further comment on this fact in appendix \ref{Einsteinsec}.}:
\be
S_{flavor}=-{ T_5 N_f \over 4\pi}
\int d^8x \frac{\sqrt{-g_{(8)}}}{e^{2\nu}}
e^{-\phi}\,\,.
\label{flavact}
\ee

By defining, for notational convenience:
\be
Q_f={ \kappa_{(8)}^2 T_5 \over 2 \pi} N_f\,\,,
\label{Qfdefinition}
\ee
and inserting the ansatz (\ref{8dmetric}), (\ref{RRF5}), (\ref{flavact})
in the action (\ref{d=8 action}) (with the usual change of sign for 
the $F_{(5)}^2$ term), we obtain the effective action which will
be the starting point of the analysis in the following sections:
\bear\label{Leff}
L_{eff}&=&e^{4\lambda+2\nu+f-2\phi}[12 \dot\lambda^2 +2 \dot\nu^2
+16 \dot\lambda\dot\nu+8\dot f\dot\lambda+4\dot f\dot\nu
-4 \dot f\dot\phi
-16 \dot\lambda \dot\phi -8 \dot\nu\dot\phi\rc
&&+4 \dot\phi^2 +2 -\ha e^{2f-4\nu}+2 e^{-2\nu}
-Q_c^2 e^{2\phi-4\nu-2f}-Q_f \,e^{\phi-2\nu}]\,\,.
\eear
From now on, a dot stands for the derivative with respect to $\tau$.
The consistency of this effective action approach
 is checked in appendix \ref{Einsteinsec},
where it is shown that the solutions found from this lagrangian satisfy
the whole set of Einstein equations.

For completeness, we write here the expressions for
the curvature scalar and determinant of the metric (\ref{8dmetric})
used to derive (\ref{Leff}):
\bear
&&R=-\ha e^{2f-4\nu} +2 e^{-2\nu} -2 \dot f^2
-20\dot\la^2-16\dot\la \dot\nu -6\dot\nu^2
-8\dot f \dot\la -4\dot f\dot \nu -2{\ddot f}-8\ddot\la-4\ddot\nu\ ,\rc
&&\sqrt{-g_{(8)}}=\sin\theta\ e^{4\la +2\nu +f}\,.
\eear

\subsection{General $AdS_5\times \tilde S^3$ solutions}
\label{secgeneral}

Since we are interested in the string duals of conformal 
field theories,
the most natural ansatz has an $AdS$ form:
\be
\lambda=k_0\tau\spa\qquad\qquad \phi,\,\nu\,,f\,=\, {\rm constants}\,,
\ee
where the constant $k_0$ is the inverse of the $AdS$ radius:
\be
R_{AdS}=\frac{1}{ k_0}\,\,.
\ee
By inserting this ansatz in the equations of motion derived from
(\ref{Leff}), one finds a 
two-parameter (depending on the $\frac{Q_c}{ Q_f}$ ratio
and also on $Q_f$ through the dilaton) family of solutions:
\be
ds_8^2=e^{2k_0\tau}\,dx_{1,3}^2+d\tau^2+
e^{2\nu_0}(d\theta^2 + \sin^2\theta\, d\varphi^2)+e^{2f_0}
(d\psi+\cos\theta\, d\varphi)^2\,\,,
\label{8dmetric2}
\ee
with:
\be\label{AdS repar}
e^{\phi_0}=\frac{2}{3}\frac{A}{Q_f}\ ,\qquad
e^{2f_0}=\frac{1}{2}(1-\frac{A}{3})(A+1)\ ,
\qquad e^{2\nu_0}=\frac{1}{2}(A+1)\ ,\qquad
 k_0^2=\frac{A+3}{12(A+1)}\,\,,
\ee
and the parameter $A$  a root of the following cubic equation:
\be\label{cubic A}
A^3+\left(\frac{16}{3}\,\frac{Q_c^2}{Q_f^2}-1\right)A^2-5 A-3=0\,\,.
\ee
This equation has three real roots (one positive and two negative ones) 
for every value of $\frac{Q_c }{ Q_f}$.
Notice, however, that, in order to give a sensible solution
(\ref{AdS repar}), the only allowed  values of $A$ lie in the range 
$0\leq A\leq 3$. Notably,
the positive root of (\ref{cubic A}) is a monotonic
function of the ratio $\frac{Q_c }{ Q_f}$ and
it is easy to check that
$A\to 3$ when $\frac{Q_c }{ Q_f}\to 0$ and that $A\to 0$ when
$\frac{Q_c }{ Q_f}\to \infty$. 
Therefore, every value of $\frac{Q_c }{ Q_f}$ yields a single
$AdS_5 \times \tilde S^3$ metric and the
 squashing of the Berger sphere is a monotonic
function of this ratio.
  In the flavorless limit $A\to 0$, $f_0=\nu_0$
and the three-sphere becomes round; this is precisely the
$AdS_5 \times S^3$ solution found in \cite{KupSon}.

As generally expected for 
non-critical strings,
all these solutions have string-scale curvatures and,
unlike the usual critical cases, there is no limit 
in which they become weakly curved. Therefore, one should take into
account higher order string corrections to the action. 
Nevertheless, as argued in
\cite{Poly,KupSon,KlebMald}, one can expect that the high symmetry of the
$AdS$ configurations (\ref{8dmetric2})
protects the form of the solutions, and that string
corrections will, at most, change the radii by some factor of order one.

On the other hand, the dilaton is parametrically small, 
so the solutions do not have string loop corrections, as usual.

\subsection{A special solution}\label{sect: special solution}

The solutions displayed in section
\ref{secgeneral} were constructed without making any reference to
the supersymmetric vacuum (\ref{8dvacuum}), (\ref{n2metric}).
Actually, there is a very special solution among this family,
which can be shown to be the result of 
 piling up branes on the $\mathbb{R}^{1,3}\times\,$\kkl vacuum (see section \ref{sect BPS}).
This solution  corresponds to taking:
\be
\frac{Q_f}{ Q_c}=\sqrt{\frac{2}{ 3}}\,\,.
\label{NfoverN}
\ee
Plugging (\ref{NfoverN}) in (\ref{cubic A}), we find $A=1$, so
the gravity solution is 
given by (\ref{8dmetric2}) and (\ref{RRF5}), where the dilaton
and the coefficients in  the metric  are (\ref{AdS repar}):
\be
e^{\phi_0}=\frac{2 }{ 3 Q_f}\,\,,\qquad
e^{2f_0}=\frac{2 }{ 3}\,\,,\qquad
e^{2\nu_0}=1\,\,,\qquad
k_0^2=\frac{1}{6}\,\,.
\label{confpoint}
\ee

As it will be shown in section \ref{subsect: superpo}, this special configuration
 is the only one among (\ref{8dmetric2})-(\ref{cubic A})
for which one can write a simple, explicit superpotential such that
the solution satisfies the corresponding system of
first-order BPS equations.   In section~\ref{sect: flow} we will also build a numerical solution interpolating between this special $AdS$ solution (\ref{confpoint}) in the IR and the $\mathbb{R}^{1,3}\times$ \kkl vacuum (\ref{8dvacuum}) in the UV.
These facts support the conjecture that this solution  is supersymmetric. For all other solutions, on the contrary, similar arguments cannot be made. 

As we show in appendix \ref{app: exist W}
using a general argument of \cite{KupSon}, 
we expect all solutions (\ref{8dmetric2})-(\ref{cubic A})
 to arise from a superpotential, but its form to be in general too complicated to be obtained analytically.
 Being the solution with $Q_f/Q_c=\sqrt{2/3}$ the only one to admit a simple superpotential, 
we argue that in some sense it is ``the most symmetric" one, its  larger symmetry constraining more strongly the system and therefore making the associated superpotential likely to turn out to be easier to write.  Here ``symmetric" of course does not refer to the isometry of the background since 
 all other solutions have the same isometry. It could well relate
to supersymmetries. A similar argument for the identification of supersymmetric vacua was also suggested in \cite{KT}.

We conjecture, therefore, that the special solution (\ref{confpoint}) is dual 
 to an ${\cal N}=1$ superconformal theory. This theory should be such that  at 
 the IR fixed point one particular ratio  $N_f/N_c$ is singled out among all 
 other possibilities, as signaled by (\ref{NfoverN}).

On general grounds, we expect that each of the solutions 
in (\ref{AdS repar})
is dual to a distinct  conformal gauge theory with 
$SU(2)\times U(1)$ global symmetry.
We may  however conclude, following \cite{KupSon}, that either  
all other
 solutions are not stable, 
 or they correspond to  a family
of unknown, possibly non-supersymmetric, conformal field theories with the
aforementioned global symmetry. Notice that the existence of a superpotential, by itself, does not guarantee that a solution  is supersymmetric~\cite{nonsusy}.

In view of the field theory interpretation of the special solution (\ref{NfoverN}), it is of evident interest to know what the relation (\ref{NfoverN}) implies in terms of the actual number of D3 and D5 branes. Let us assume the
usual quantization condition for the D3 brane charge:
\be
{1 \over 2\kappa_{(8)}^2} \int_{\tilde S^3}
{}^*F_{(5)}\,=\,N_c \, T_3 \,=\, {1 \over 2\kappa_{(8)}^2} 16 \pi^2 Q_c\,\,.
\label{quantNc}
\ee
For the first equality, we have assumed that,
as in critical string theory, the charge and tension
of the brane are equal, while the second equality is obtained by
direct computation with the solution. From (\ref{Qfdefinition}),
(\ref{NfoverN}) and (\ref{quantNc}), we then obtain:
\be
{N_f \over N_c}\,=\,\pi\,\sqrt{2\over 3}\simeq 2.56\,\,.
\label{NfNratio}
\ee
Notice that to derive this expression the ratio of tensions
${T_3 \over T_5}=(2\pi)^2$ has been used, but the precise
expressions for $T_3$, $T_5$, $\kappa_{(8)}$ were not necessary.

Clearly, the above irrational value for ${N_f \over N_c}$ cannot
 match  the actual field theory value, which by definition 
 has to be rational. 
 One obvious possibility is that stringy corrections here are of order one.
 For instance, terms with
higher powers of $F_{(5)}$ in the lagrangian would affect the relation
between $Q_c$ and $N_c$, and corrections to the DBI action would alter
the value of ${Q_f \over N_f}$. Furthermore, all sorts of corrections,
like higher curvature terms, may in part modify the superpotential 
calculation
of section \ref{sect BPS}, possibly leading to a value of ${Q_f \over Q_c}$
different from (\ref{NfoverN}).
Finally, another possibility is that the normalization of the brane tensions
 we are using is incorrect in this type of background.

\subsection{Spectrum of perturbations around $AdS_5\times$ $\tilde{S}^3$  and stability}\label{subsect: stability}
We consider now the spectrum of perturbations around the solutions of the form
 $AdS_5\times \tilde{S}^3$ we found in section \ref{secgeneral}. 
 This is interesting mainly for two reasons: first of all, it will
 test the stability\footnote{We do not discuss the closed string tachyon since we lack a satisfactory description of this mode. We expect, as in \cite{KlebMald}, that it will not cause any instabilities on our backgrounds.} of these backgrounds against the excitation of some metric and dilaton modes. Secondly, it will be essential for the construction of  a flow connecting the special $AdS$ solution of section \ref{sect: special solution} to the $\mathbb{R}^{1,3}\times$ \kkl vacuum, as we will see in section \ref{sect BPS}.

Let us start from the metric (\ref{8dmetric}) and effective action (\ref{Leff}) in the string frame. It is useful to define a new radial coordinate $u$ and new fields $\alpha$, $\beta$, $\gamma$:
\be\label{convenient}
\begin{split}
&d\tau=e^{\frac{2}{3}\phi-\frac{2}{3}\nu-\frac{1}{3}f}\;du\ ,\\
&\alpha=-\frac{2}{3}\phi+\frac{2}{3}\nu+\lambda+\frac{1}{3}f\ ,\\
&\beta=-\frac{1}{3}\phi+\frac{2}{3}\nu+\frac{1}{3}f\ ,\\
&\gamma=\frac{1}{3}\nu-\frac{1}{3}f\ ,
\end{split}
\ee
in terms of which the metric reads:
\be\label{metricu}
ds^2=e^{\frac{2}{3}\phi-2\beta}(du^2+e^{2\alpha}\,dx^2_{1,3})
+e^{\frac{2}{3}\phi+2\beta+2\gamma}(d\theta^2+\sin^2\theta\,d\varphi^2)
+e^{\frac{2}{3}\phi+2\beta-4\gamma}(d\psi+\cos\theta\,d\varphi)^2\ .
\ee
The field $\gamma(u)$ determines the squashing of the three-sphere: when $\gamma=0$ the three-sphere is perfectly round. The effective action becomes:
\be\label{Seffu}
S_{eff}=4\int du\;e^{4\alpha}\left(3\alpha'^{\,2}-\frac{1}{6}\phi'^{\,2}-\frac{3}{2}\beta'^{\,2}-\frac{3}{2}\gamma'^{\,2}-V(\phi,\beta,\gamma)\right)\ ,
\ee
where $'$ denotes derivation with respect to $u$, and the potential $V$ does not depend on $\alpha(u)$:
\be\label{Veffu}
V(\phi,\beta,\gamma)=-\frac{1}{2}e^{\frac{2}{3}\phi-2\beta}+\frac{1}{8}e^{-4\beta-8\gamma}-\frac{1}{2}e^{-4\beta-2\gamma}+\frac{1}{4}Q_c^2e^{\frac{2}{3}\phi-8\beta}+\frac{1}{4}Q_fe^{\phi-4\beta-2\gamma}\ .
\ee

It is straightforward to write the equations of motion for these new fields in terms of the radial coordinate $u$:
\be\label{eom pabc}
\begin{split}
&3\alpha''+6\alpha'^{\,2}+\frac{1}{3}\phi'^{\,2}+3\beta'^{\,2}+3\gamma'^{\,2}+2V=0\ ,\\
&\phi''+4\alpha'\phi'=3\frac{\partial V}{\partial \phi}\ ,\\
&\beta''+4\alpha'\beta'=\frac{1}{3}\frac{\partial V}{\partial \beta}\ ,\\
&\gamma''+4\alpha'\gamma'=\frac{1}{3}\frac{\partial V}{\partial \gamma}\ .
\end{split}
\ee
These equations are  satisfied by our $AdS_5\times \tilde{S}^3$ solutions (\ref{AdS repar}) which, in terms of the new variables, read:
\begin{align}\label{solut beta}
e^{\phi_0}&=\frac{2}{3}\frac{A}{Q_f}\ ,& e^{\beta_0}&=Q_f^{\frac{1}{3}}\left(\frac{3}{32}\frac{(3-A)(A+1)^3}{A^2}\right)^\frac{1}{6}\ ,\\
e^{\gamma_0}&=\left(\frac{3}{3-A}\right)^\frac{1}{6}\ , &\alpha&=\alpha_0+\tilde{k}_0 u\ ,\nonumber
\end{align}
where $\alpha_0=-\frac{1}{3}\phi_0+\beta_0$, $\tilde{k}_0$ is related to $k_0$ in (\ref{AdS repar}) through $\tilde{k}_0=e^{\phi_0/3-\beta_0}k_0$ and, as before, $A$ depends on the ratio $Q_c/Q_f$ through the cubic equation (\ref{cubic A}).

The masses of (a subset of) the fluctuations around these solutions, are the eigenvalues of the laplacian acting on the small perturbations $\delta\phi$, $\delta\beta$ and $\delta\gamma$:
\be\label{flucts}
\begin{split}
&\phi(u)=\phi_0+\delta\phi(u)\ ,\\
&\beta(u)=\beta_0+\delta\beta(u)\ ,\\
&\gamma(u)=\gamma_0+\delta\gamma(u)\ ,
\end{split}
\ee
with $\alpha(u)$ unperturbed\footnote{This choice guarantees that the first equation in (\ref{eom pabc}) is automatically satisfied at first order in the perturbations.}. In our specific case, the laplacian reads:
\be
\nabla^2 f(u)=\frac{1}{\sqrt{-g}}\partial_\mu(\sqrt{-g}\,g^{\mu\nu}\partial_\nu f(u))=e^{-\frac{2}{3}\phi+2\beta}(f''+2\phi'f'+4\alpha'f')\ ,
\ee
and by substituting into this expression the equations of motion (\ref{eom pabc}), we find, to first order in the perturbations,
\be\label{laplacian}
\begin{split}
&\nabla^2(\delta\phi)=3e^{-\frac{2}{3}\phi_0+2\beta_0}\left( \left.\frac{\partial^2V}{\partial \phi^2} \right|_0\delta\phi + \left.\frac{\partial^2V}{\partial \beta\partial \phi} \right|_0\delta\beta+ \left.\frac{\partial^2V}{\partial \gamma\partial \phi} \right|_0\delta\gamma\right)\ ,\\[5pt]
&\nabla^2(\delta\beta)=\frac{1}{3}e^{-\frac{2}{3}\phi_0+2\beta_0}\left( \left.\frac{\partial^2V}{\partial \phi\partial \beta} \right|_0\delta\phi + \left.\frac{\partial^2V}{\partial \beta^2} \right|_0\delta\beta+ \left.\frac{\partial^2V}{\partial \gamma\partial \beta} \right|_0\delta\gamma\right)\ ,\\[5pt]
&\nabla^2(\delta\gamma)=\frac{1}{3}e^{-\frac{2}{3}\phi_0+2\beta_0}\left( \left.\frac{\partial^2V}{\partial \phi\partial \gamma} \right|_0\delta\phi + \left.\frac{\partial^2V}{\partial \beta\partial \gamma} \right|_0\delta\beta+ \left.\frac{\partial^2V}{\partial \gamma^2} \right|_0\delta\gamma\right)\ .
\end{split}
\ee

We were not able to find a simple explicit form for the eigenvalues of $\nabla^2$,
but we can at least evaluate them numerically. One thing can be noticed right
away: the explicit dependence of the masses of the fluctuations on $Q_f$ or $Q_c$ cancels, and the spectrum only depends on their ratio $Q_f/Q_c$ through the $A$-dependence\footnote{This is different
from what happens for the duals of $\mathcal{N}=1$ superconformal QCD
\cite{KlebMald}, where the masses of the fluctuations around the $AdS$ solutions
are completely independent of $N_c$ and $N_f$.}. For this reason, solutions corresponding to different values of $Q_f/Q_c$ might have very different behavior. We plot in figure \ref{fig m2} the graph of the squared masses of the three fluctuations we consider, as functions of $A$.

\FIGURE[!ht]{\epsfig{file=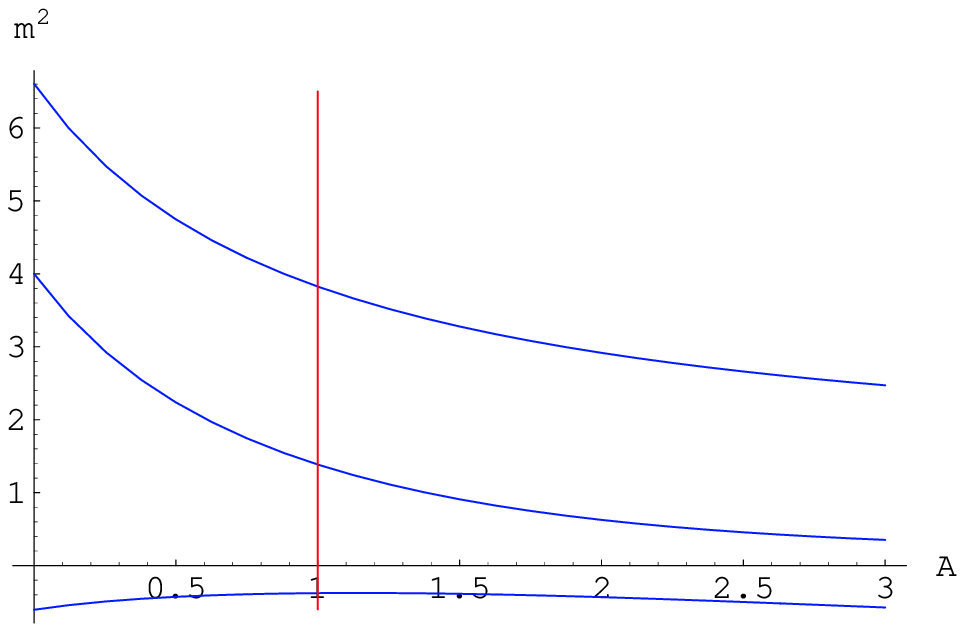,width=0.8\textwidth}
\caption{Squared masses of fluctuations around the $AdS$ solutions. 
Notice that one of the eigenvalues is negative for all allowed values of $A$: $0\leq A\leq 3$. 
The vertical line represents the special value $A=1$.}
\label{fig m2}}
Notice that one of the three squared masses is always negative.
 Since the background we are considering is $AdS$,
  we have to ensure that this eigenvalue satisfies the 
Breitenlohner-Freedman (BF) bound in order for the solution to be stable \cite{BF}:
\be\label{BF bound}
m^2\geq -\frac{4}{R_{AdS}^2}\ ,
\ee
where, in our case, $R_{AdS}^2=\frac{1}{k_0^2}=\frac{12(A+1)}{A+3}$. The BF stability condition is checked in figure~\ref{fig m2low + BF}, where the solid  curve represents the squared mass of the lightest mode, and the dashed curve the extremal  BF bound.
\FIGURE[!ht]{\epsfig{file=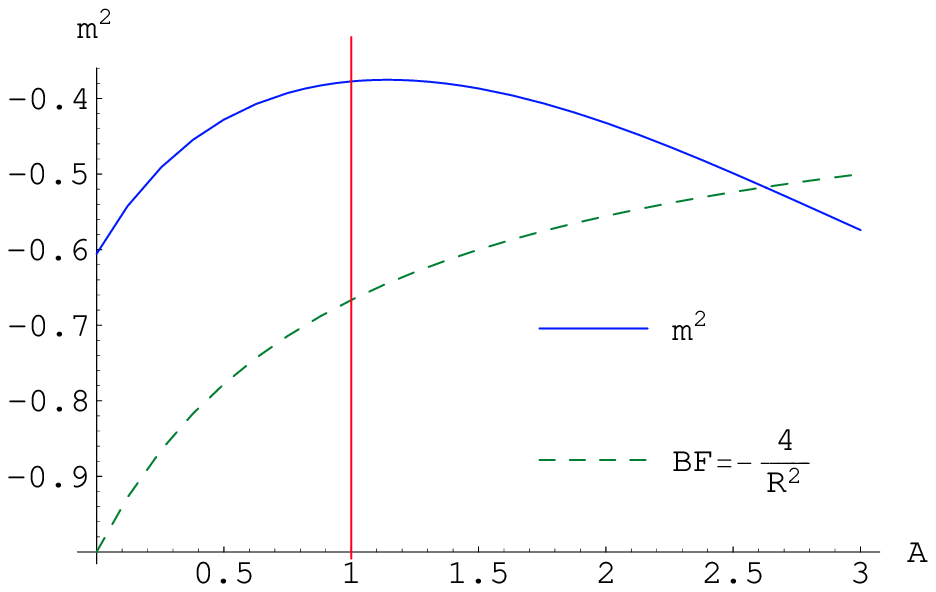,width=0.8\textwidth}
\caption{Stability of the solutions against the Breitenlohner-Freedman bound. The solid curve represents the squared mass of the lightest mode of fluctuations around the $AdS$ solutions while the dashed line is the BF boundary. We notice that the bound is not satisfied for all allowed values of $A$. Again the vertical line shows the special case $A=1$.}
\label{fig m2low + BF}}

We notice that not all of our solutions satisfy  (\ref{BF bound}): the BF bound is violated for all values $A>A_c$, where:
\be
A_c=2.625...
\ee
which corresponds to:
\be\label{max ratio}
\frac{Q_f}{Q_c}=2.729...
\ee
As we pointed out after (\ref{cubic A}), $A$ is a monotonic function of $Q_f/Q_c$, therefore all backgrounds with $Q_f/Q_c$ larger than the value (\ref{max ratio}) are unstable.

\subsection*{Fluctuation spectrum of the special solution $A=1$}

Since  the special solution of section \ref{sect: special solution} corresponding to $Q_f/Q_c=\sqrt{2/3}$ (and $A=1$) plays an important role, we focus here on this case. For the eigenvalues of (\ref{laplacian}) we find:
\be\label{statemasses}
m^2_1=-0.377... \qquad \qquad m^2_2=1.385...\qquad\qquad m^2_3=3.825...
\ee
On the gauge theory side, a scalar fluctuation of mass $m$ around a string background corresponds to a chiral operator with dimension\footnote{For states with $m^2R_{AdS}^2>-3$ the only consistent choice of sign in the operator dimension formula is~$+$~\cite{KlebWitt}.} $\Delta=2\pm\sqrt{4+m^2 R_{AdS}^2}$ \cite{GKP,W}. As we argued above,  our solutions are subject to string corrections even at leading order, because the curvature of the background is always of the same order of magnitude as the string scale; therefore the spectrum (\ref{statemasses}) will receive corrections. Nonetheless, expecting that the structure of our $AdS$ solutions will not be affected by string corrections, we  
anticipate that the higher order contributions we have neglected, will not correct drastically our results. Then, we can give a rough estimate of the conformal dimension of the operators dual to the fluctuations (\ref{flucts}) around the $Q_f/Q_c=\sqrt{2/3}$ solution by using (\ref{statemasses}); we find ($R_{AdS}^2=6$):
\be\label{confdimensions}
\Delta_1=3.317\ldots\qquad\Delta_2=5.509\ldots\qquad\Delta_3=7.191\ldots
\ee

Regarding  string corrections, it is interesting to notice that the Laplacian eigenvector corresponding to the nearly vanishing eigenvalue  $m_1^2$ is ``dilaton dominated":
\be\label{massless?}
\delta\phi+(0.120\ldots )\delta\beta+(0.065\ldots)\delta\gamma\sim \delta\phi+(0.106\ldots)\delta\nu+(0.019\ldots)\delta f\ .
\ee
In ordinary critical $AdS$/CFT the dilaton is massless and it is dual to the marginal (i.e. $\Delta=4$) operator $\mathrm{Tr}\,F^2$ of the conformal gauge theory. In analogy to this, it is tempting to guess that string corrections will make the perturbation (\ref{massless?}) massless, and therefore to identify this field with the field theory modulus associated to some marginal operator.

\section{Superpotential, BPS equations and a flow from $AdS$ to \kkl}\label{sect BPS}

The issue of conserved supersymmetries for  the solutions we have found in  section \ref{secgeneral} is not at all trivial. Unfortunately, we lack a complete supersymmetric formulation of non-critical string theory on a background with RR fluxes, and in particular we do not know what the Killing spinor equations are. As we anticipated in section \ref{sect: special solution}, in this section we will present the two main arguments which strongly suggest that the special solution $Q_f/Q_c=\sqrt{2/3}$ ($A=1$)  is supersymmetric: we will show that this is the only solution for which it is possible to write an explicit, simple superpotential, and we will find a numerical solution which confirms that the background with $A=1$ is the back-reacted solution obtained by placing D3 and D5 branes on the $\mathbb{R}^{1,3}\times$ \kkl vacuum.

\subsection{Superpotential}\label{subsect: superpo}

A necessary condition for supersymmetry is the existence of a superpotential \cite{Freedman:1999gp, nonsusy} whose related BPS equations are solved by the solutions we consider; studying this object will provide some important information about our solutions.

Let us start from the general case and suppose that the action of a system reads:
\be\label{Sgeneral}
S=\int\,du \,e^{4\alpha}\left(3(\alpha')^2-\ha G_{ab}(f)f'^af'^b-V(f)\right)\ .
\ee
The associated superpotential has to solve the following partial differential equation:
\be\label{Wequation}
V=\frac{1}{8}G^{ab}\frac{\partial W}{\partial f^a}\frac{\partial W}{\partial f^b}-\frac{1}{3}W^2\ .
\ee
When a solution to this equation does exist, the configurations satisfying  the following set of  first order BPS equations:
\be\label{BPSgeneral}
f'^a=\ha G^{ab}\frac{\partial W}{\partial f^b}\ ,\qquad\qquad \alpha'=-\frac{1}{3}W(f)\ ,
\ee
are solutions to the second order differential equations coming from  (\ref{Sgeneral}) \cite{Freedman:1999gp, nonsusy}. Therefore  finding a superpotential simplifies considerably  the problem of building solutions.

Let us focus now on the particular case we are interested in, the solutions of the form (\ref{8dmetric}), (\ref{RRF5}) to the effective action (\ref{d=8 action}). By using the same redefinitions (\ref{convenient}) as in section~\ref{subsect: stability}, we see that  (\ref{Seffu}) is in the form (\ref{Sgeneral}),
with (\ref{Veffu}):
\be\label{Veffubis}
V(\phi,\beta,\gamma)=-\frac{1}{2}e^{\frac{2}{3}\phi-2\beta}+\frac{1}{8}e^{-4\beta-8\gamma}-\frac{1}{2}e^{-4\beta-2\gamma}+\frac{1}{4}Q_c^2e^{\frac{2}{3}\phi-8\beta}+\frac{1}{4}Q_fe^{\phi-4\beta-2\gamma}\ ,
\ee
and  (disregarding an overall factor):
\be
G_{\phi\phi}=\frac{1}{3}\ ,\qquad G_{\beta\beta}=3\ ,\qquad G_{\gamma\gamma}=3\ ,
\ee
and all other components  vanishing. Equation (\ref{Wequation}) reads then:
\be\label{diffW}
\frac{3}{8}\left(\frac{\partial W}{\partial \phi}\right)^2 +\frac{1}{24}\left(\frac{\partial W}{\partial \beta}\right)^2 +\frac{1}{24}\left(\frac{\partial W}{\partial \gamma}\right)^2-\frac{1}{3}W^2=V(\phi,\beta,\gamma)\ .
\ee

We start solving this equation for the vacuum and therefore  fix $Q_c=Q_f=0$; it can be   shown then that the superpotential:
\be\label{vacuum super}
W=-\frac{1}{2}e^{\frac{2}{3}\phi-2\gamma}-e^{-2\beta+2\gamma}-\frac{1}{2}e^{-2\beta-4\gamma}
\ee 
is a solution to (\ref{diffW}). We show in appendix \ref{app: KKL from super} that the $\mathbb{R}^{1,3}\times$ \kkl vacuum (\ref{n2metric}), is the solution to the first order equations of motion we get from this superpotential.

Let us now turn on the RR flux and the flavor brane term.  Remarkably, as
anticipated in section \ref{sect: special solution}, there is
only one value of $Q_f/Q_c$ for which 
the solution  
has a very simple form.  This happens for:
\be\label{spec ratio}
\frac{Q_f}{Q_c}=\sqrt{\frac{2}{3}}\ ,
\ee
in which case the superpotential is given by\footnote{Since equation (\ref{diffW}) depends on $W$ quadratically, the overall sign of the superpotential is not determined. We fix it by requiring that the $AdS$ critical point is a local minimum of $W$ rather than a maximum.}:
\be\label{Wexact}
W=-\frac{1}{2}e^{\frac{2}{3}\phi-2\gamma}-e^{-2\beta+2\gamma}-\frac{1}{2}e^{-2\beta-4\gamma}+Q_f \,e^{\frac{1}{3}\phi-4\beta}\ .
\ee

A very important feature of our solution appears here: we can find an explicit solution to (\ref{diffW}) only for a single value of $Q_f/Q_c$. Indeed it is in general really hard, if possible, to find a superpotential that solves equation (\ref{diffW}); for this reason, the result we find is quite remarkable, and very promising for the interpretation of our solutions.

As we argued before, the dual field theory to the special solution (\ref{spec
ratio}) should be an $\mathcal{N}=1$ SCFT where the requirements for conformal
invariance and $SU(2)$ global symmetry are satisfied only for a special value of $N_f/N_c$.

\subsection*{BPS first order equations for $Q_f/Q_c=\sqrt{2/3}$}

With the superpotential (\ref{Wexact}) we found for the case $Q_f/Q_c=\sqrt{2/3}$, the first order equations of motion (\ref{BPSgeneral}) read:
\be\label{supereqn}
\begin{split}
&\partial_u \alpha=\frac{1}{6}e^{\frac{2}{3}\phi-2\gamma}+\frac{1}{3}e^{-2\beta+2\gamma}+\frac{1}{6}e^{-2\beta-4\gamma}-\frac{1 }{3}Q_f\,e^{\frac{1}{3}\phi-4\beta}\ ,\\[4pt]
&\partial_u \phi=-\frac{1}{2}e^{\frac{2}{3}\phi-2\gamma}+\frac{1}{2}Q_f e^{\frac{1}{3}\phi-4\beta}\ ,\\[4pt]
&\partial_u \beta=\frac{1}{3}e^{-2\beta+2\gamma}+\frac{1}{6}e^{-2\beta-4\gamma}-\frac{2}{3}Q_f e^{\frac{1}{3}\phi-4\beta}\ ,\\[4pt]
&\partial_u \gamma=\frac{1}{6}e^{\frac{2}{3}\phi-2\gamma}-\frac{1}{3}e^{-2\beta+2\gamma}+\frac{1}{3}e^{-2\beta-4\gamma}\ .
\end{split}
\ee
Unfortunately, we were not able to find the general solution to these equations. Nonetheless, it is straightforward to show that the  $AdS_5\times \tilde{S}^3$ solution with $Q_f/Q_c=\sqrt{2/3}$ (\ref{confpoint}) satisfies~(\ref{supereqn}). The constant values of the fields read:
\be\label{BPSsolut}
e^{\phi_0}=\frac{2}{3Q_f}\ ,\qquad e^{2\beta_0}=Q_f^{2/3}\left(\frac{3}{2}\right)^{1/3}\ ,\qquad e^{2\gamma_0}=\left(\frac{3}{2}\right)^{1/3}\ .
\ee

\subsection{A numerical flow from $AdS_5\times \tilde{S}^3$ to the $\mathbb{R}^{1,3}\times$ \kkl vacuum} \label{sect: flow}

In this section we study the solution of the first order BPS equations (\ref{supereqn}) obtained by perturbing the special solution (\ref{BPSsolut}). As we will see, this will give rise to a domain wall  interpolating between two different vacua of the eight-dimensional non-critical string theory: our solution $Q_f/Q_c=\sqrt{2/3}$ in the small radius region (IR) and the $\mathbb{R}^{1,3}\times$ \kkl vacuum in the large radius region (UV).

Before presenting the calculation, we will discuss  what the interpretation of the flow we find should be.
 The domain wall represents the back-reaction of the stack of D3 and D5-branes we described in section \ref{setup}, on the whole space-time and not just in the near-horizon limit. When we are in the IR region close to the D3-branes, the D5-branes are densely distributed on the transverse two-sphere and the background is our solution $AdS_5\times \tilde{S}^3$, but as we go further away from this region the two-sphere grows and the D5-branes dilute. When we reach the large $\tau$ region (UV), the D3-branes are very far away and the density of the D5's is almost zero: the space-time feels a vanishing influence from the brane system and therefore it approaches the (empty) vacuum. Remarkably, in the UV region we find that the background asymptotes  the $\mathbb{R}^{1,3}\times$ \kkl vacuum. This confirms that the solution with $Q_f/Q_c=\sqrt{2/3}$ is the  near-horizon limit of the background generated by a stack of D3 and D5-branes placed on the \kkl vacuum.

Let us now construct the flow solution. For the following analysis, it is convenient to switch back to the radial coordinate $\tau$. This amounts to the following redefinition of derivatives:
\be
\dot{x}\equiv\partial_\tau x =e^{-\frac{1}{3}\phi+\beta}\partial_u x\ .
\ee
Since the field $\alpha(\tau)$ does not enter the equations (\ref{supereqn}) for the  fields $\phi$, $\beta$ and $\gamma$, which are those we are interested in to match with the fluctuations of section \ref{sect: special solution}, we will leave it aside for the moment and concentrate on:
\be
\phi(\tau)=\phi_0+\delta\phi(\tau)\ ,\qquad \beta(\tau)=\beta_0+\delta\beta(\tau)\ ,\qquad\gamma(\tau)=\gamma_0+\delta\gamma(\tau)\ .
\ee
We can expand the BPS equations (\ref{supereqn}) up to first order in the perturbations, and obtain:
\be\label{1st ord mat}
\partial_\tau\left(\begin{array}{c} \delta\phi\\[4pt] \delta\beta\\[4pt]\delta\gamma\end{array}\right)=\frac{1}{\sqrt{6}}\left(\begin{array}{rrr}-\frac{1}{3} & -4 & 2\\[4pt]
-\frac{4}{9} & \frac{8}{3} & \frac{2}{3}\\[4pt] \frac{2}{9} & \frac{2}{3} & -\frac{16}{3}\end{array}\right)\left(\begin{array}{c} \delta\phi\\[4pt] \delta\beta\\[4pt] \delta\gamma\end{array}\right)\ .
\ee
The eigenvalues of this matrix are given by:
\be\label{eigenvalues}
\frac{1}{\sqrt{6}}(\lambda_1,\lambda_2,\lambda_3)=\frac{1}{\sqrt{6}}(-0.683\ldots, -5.509\ldots, 3.191\ldots)\ .
\ee
The comparison of these eigenvalues to the conformal dimensions (\ref{confdimensions}) will tell us whether on the dual field theory side these deformations  amount to adding an operator or turning on a non-zero expectation value \cite{GKP,W}. To do so, we need to introduce the coordinate $z=\sqrt{6}e^{-\frac{\tau}{\sqrt{6}}}$, with respect to which the $AdS$ metric reads $ds^2=6(dz^2+dx_{1,3}^2)/z^2$. The first and second perturbations grow in the IR $(z\gg 1)$ as $z^{-\lambda_1}=z^{4-\Delta_1}$ and  $z^{-\lambda_2}=z^{\Delta_2}$ respectively, and therefore the first corresponds to adding a relevant  (or marginal, see the comment at the end of section \ref{subsect: stability}) operator to the action of the field theory, while the second one corresponds to giving an expectation value to an operator in the gauge theory. The  third perturbation, on the contrary, grows in the UV $(z\sim 0)$ as  $z^{-\lambda_3}=z^{4-\Delta_3}$  and corresponds to adding an irrelevant operator to the action. Since we are interested in deforming the theory in the IR with an irrelevant operator, this last irrelevant, \mbox{UV-growing}  deformation is the one we will turn on: we start in the IR with a configuration which asymptotes our $AdS$ solution $(\phi_0,\beta_0,\gamma_0)$, and deform it with the  eigenvector of (\ref{1st ord mat}) corresponding to $\lambda_3$ (\ref{eigenvalues}):
\be
\phi=\phi_0+\delta\phi\ ,\qquad \beta=\beta_0-(0.903\ldots)\delta\phi\ ,\qquad\gamma=\gamma_0-(0.045\ldots)\delta\phi\ ,
\ee
which will drive the solution away from $AdS$ in the UV. To study this flow, and eventually compare the UV asymptotic region with the $\mathbb{R}^{1,3}\times$ \kkl vacuum, it is convenient to analyze how the combinations of fields appearing in the metric (\ref{metricu}) evolve, rather than the fields $\alpha$, $\beta$ and $\gamma$, that is to study the flow of:
\be\label{repar flow}
\lambda=\frac{1}{3}\phi-\beta+\alpha\ ,\qquad\nu = \frac{1}{3}\phi+\beta+\gamma\ ,\qquad f=\frac{1}{3}\phi+\beta-2\gamma\ ,
\ee
and the dilaton $\phi$. To compare the UV asymptotic behavior of the flow with the \kkl va\-cuum, we also  introduce the same radial coordinate that enters the metric (\ref{n2metric}) for the vacuum:
\be
d\zeta=4\;e^{\frac{1}{3}\phi+\beta-2\gamma}d\tau\ .
\ee
The first order equations we need to solve read then:
\be
\begin{split}
&\partial_\zeta\phi=-\frac{1}{8}+\frac{1}{8}Q_f e^{\phi-2\nu-2f}\ ,\\
&\partial_\zeta\nu=\frac{1}{8}e^{-2\nu}-\frac{1}{8}Q_f e^{\phi-2\nu-2f}\ ,\\
&\partial_\zeta f=-\frac{1}{8}+\frac{1}{4}e^{-2f}-\frac{1}{8}e^{-2\nu}-\frac{1}{8}Q_fe^{\phi-2\nu-2f}\ ,\\
&\partial_\zeta \lambda=\frac{1}{8}Q_fe^{\phi-2\nu-2f}\ ,
\end{split}
\ee
and the initial conditions we impose at $\zeta=0$ (that is in the IR region) are (\ref{confpoint}):
\be
\phi_0=\log\frac{2}{3Q_f}\ ,\qquad\nu_0=0\ ,\qquad f_0=\log\sqrt{\frac{2}{3}}\ ,\qquad\lambda(0)=0\ .
\ee

If we take $\delta\phi=0$ there is obviously no flow, and we get our $AdS$ solution on the whole space. If $\delta\phi$ is taken to be positive, the evolution of the fields blows up at some finite value of $\zeta$. If we take $\delta\phi<0$ the fields approach the \kkl vacuum value in the large $\zeta$ region. This requires the initial value of $\phi_{0,\rm KKL}$ to be tuned, which is a natural request if we think that in the vacuum the dilaton is linear and its value at $\zeta=0$ is a free parameter, whereas in the $AdS$ background the dilaton is fixed by the equations of motion. The numerical analysis is reported in figure \ref{fig: flow}.

\FIGURE[!ht]{\epsfig{file=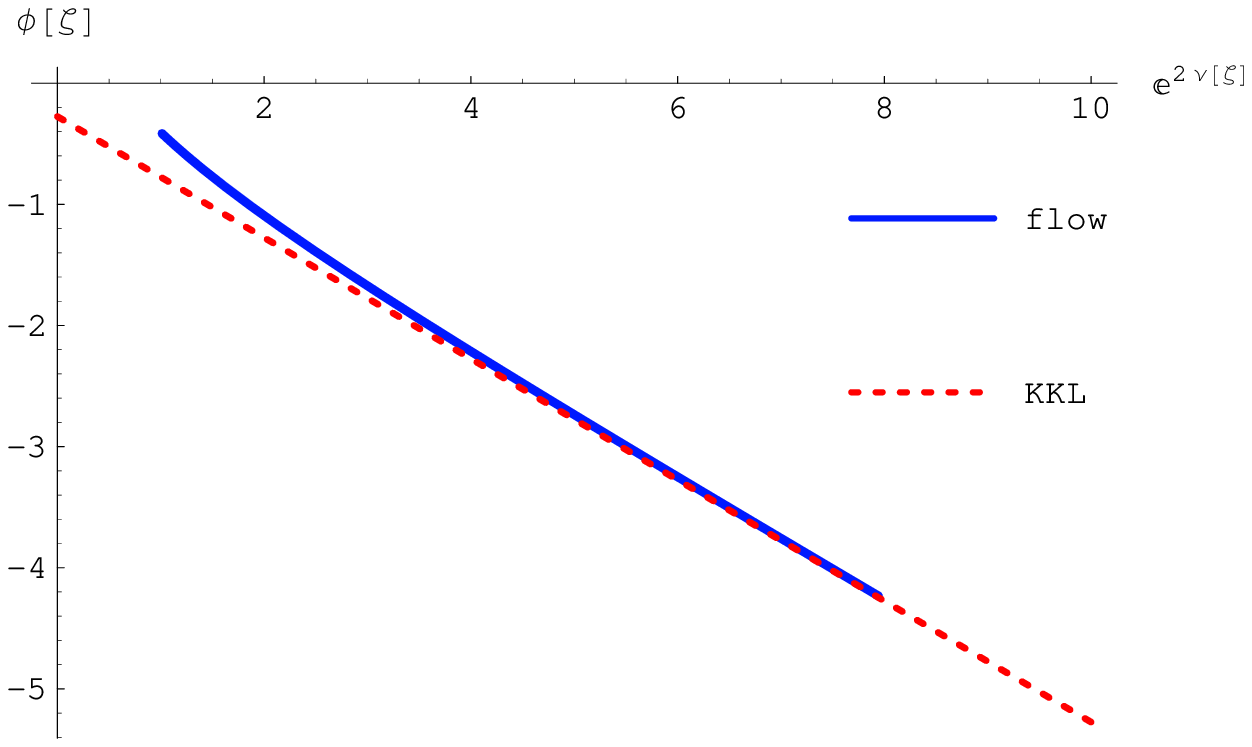,width=0.546\textwidth}\epsfig{file=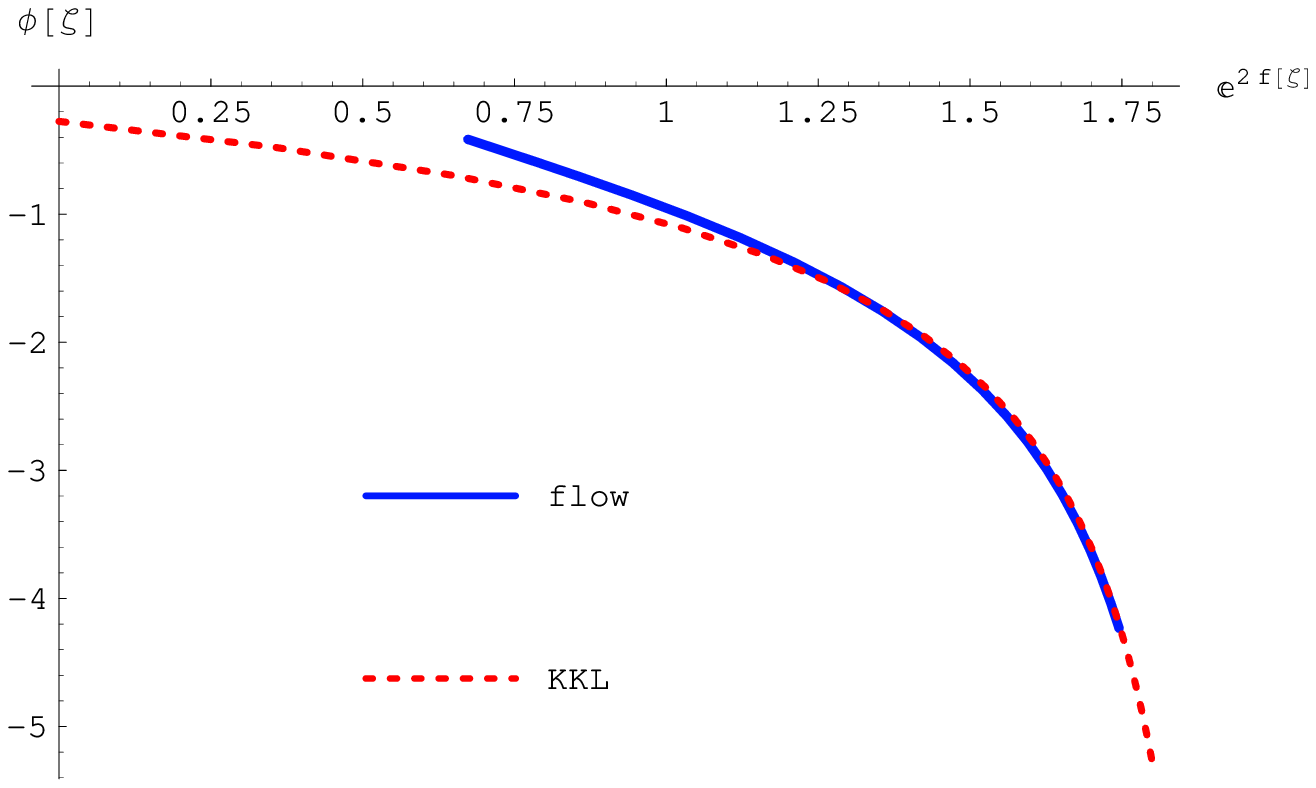,width=0.52\textwidth}
\caption{Relation between the dilaton $\phi(\zeta)$ and the fields $\nu(\zeta)$ (left) and $f(\zeta)$ (right) in the flow and in \kkl. The flow starts from the $AdS_5\times \tilde{S}^3$ solution. In both graphs the two curves overlap for decreasing $\phi$, i.e. in the UV. The thicker line represents the flow solution, while the dashed line is the \kkl vacuum.}
\label{fig: flow}}


\section{Comments on the $\mathcal{N}=1$ dual field theory}\label{sect: finter}

It would be very important to identify the dual  field theory to our special $AdS_5\times \tilde{S}^3$ solution~(\ref{confpoint}). Unfortunately, because of the intrinsic limitations 
of the gravity approximation to non-critical string theory, and a lack 
of knowledge of the low energy dynamics of branes on \kkl, we are not 
able to give a final answer to this  question.

As we argued in section \ref{setup}, we believe that the $AdS_5\times 
\tilde{S}^3$ solution (\ref{confpoint}) preserves eight supercharges 
(four background supercharges and four near-horizon conformal ones). 
This  requires the dual field 
theory to be an $\mathcal{N}=1$ superconformal gauge theory.
The theory will contain flavors in the (anti) fundamental representation, 
and possibly other matter fields. Due to the symmetry of the dual 
background, it is conceivable that some $SU(2)$ invariant superpotential 
for these fields will be present at the IR fixed point. This will also 
contribute to fix the $N_f/N_c$ ratio. 

Some hints toward an answer could come from the string engineerings of supersymmetric gauge theories and the description of NS5-brane configurations in terms of conformal field theories.  It is well known \cite{NSbranes} that an $\mathcal{N}=2$ $SU(N_c)$  gauge theory can be engineered by placing two parallel NS5-branes on flat space-time and attaching to them $N_c$ D4-branes. Flavors can be introduced by adding D6-branes to the picture with additional D4-branes stretching from either NS-brane to any D6-brane. The general configuration is summarized in table \ref{N=2 NS branes}.
\begin{table}[ht]
\begin{center}
\begin{tabular}{|c|c|c|c|c|c|c|c|c|c|c|}
\hline
\multicolumn{1}{|c|}{ }
&
\multicolumn{4}{|c|}{$x_{1,3}$}
&\multicolumn{1}{|c|}{$4$}
&\multicolumn{1}{|c|}{$5$}
&\multicolumn{1}{|c|}{$6$}
&\multicolumn{1}{|c|}{$7$}
&\multicolumn{1}{|c|}{$8$}
&\multicolumn{1}{|c|}{$9$}\\
\hline
NS5 &$-$&$-$&$-$&$-$&$-$&$-$&$\cdot$&$\cdot$&$\cdot$&$\cdot$\\
\hline
D4 &$-$&$-$&$-$&$-$&$\cdot$&$\cdot$&$-$&$\cdot$&$\cdot$&$\cdot$\\
\hline
D6 &$-$&$-$&$-$&$-$&$\cdot$&$\cdot$&$\cdot$&$-$&$-$&$-$\\
\hline
\end{tabular}
\caption{Type IIA configuration engineering $\mathcal{N}=2$ QCD.
\label{N=2 NS branes}}
\end{center}
\end{table}

A very interesting result, which is the base  for the connection to non-critical string theories, states that a system of $m$ separated NS5-branes is described by an exactly solvable CFT \cite{OV}:
\be
\left( \frac{SL(2,\mathbb{R})_{m+2}}{U(1)}\times \frac{SU(2)_{m-2}}{U(1)}\right)/{\mathbb{Z}_m}\ ,
\ee
where $\frac{SL(2,\mathbb{R})_{m+2}}{U(1)}$ is the 2d cigar
geometry, which has an asymptotically linear dilaton, while
$\frac{SU(2)_{m-2}}{U(1)}$ stands for the $\mathcal{N}=2$ minimal
model with central charge $3(1-\frac{2}{m})$. In particular, 
in the case $m=2$ we
are interested in, the central charge of the $SU(2)/U(1)$ conformal field theory is zero, and the target space directions which it describes play no role in the string theory dynamics. Attaching D-branes to the NS5-branes as in table \ref{N=2 NS branes} corresponds then to suitably placing  D3 and D5-branes in a non-critical string theory background containing the cigar.
We conjecture that considering  eight-dimensional  string theory on the vacuum:
\be\label{vac 4 N=2}
\mathbb{R}^{1,3}\times \mathbb{R}^2 \times \frac{SL(2,\mathbb{R})_{4}}{U(1)}\ ,
\ee
and placing (Minkowski-filling) D3 branes at the tip of the cigar gives rise to an $\mathcal{N}=2$ gauge theory (for similar considerations see \cite{israel}). In particular $\mathcal{N}=2$ superconformal QCD will be obtained by placing $N_c$ D3-branes at the tip of the cigar, and $2N_c$ D5-branes completely wrapping the cigar.
A first immediate check involves the Coulomb branch of the $\mathcal{N}=2$ theory: the D3-branes cannot move from the tip of the cigar, but are free to move on the transversal $\mathbb{R}^2$, describing the classical Coulomb branch of an $\mathcal{N}=2$ theory. Notice that the $SU(2)_R$ symmetry is not visible as an isometry of the vacuum \cite{Murthynotes}.

As we described in section \ref{setup}, the vacuum where we place the 
branes is different from~(\ref{vac 4 N=2}): our vacuum is the \kkl 
space of \cite{KKL}. It has been shown in \cite{HoriKap} that this 
space is related to parallel NS5-branes wrapping a vanishing 
$\mathbb{CP}^1$ inside the Eguchi-Hanson space, so
the number of preserved supercharges
 is eight. 
 The corresponding world-sheet CFT is given 
 by~\cite{HoriKap}:
\be\label{LSM1}
\left[ \frac{\mathrm{LSM}_{4m}}{\mathbb{Z}_2}\times 
\frac{SU(2)_{m-2}}{U(1)}\right]/\mathbb{Z}_m^{\mathrm{diag}}\ ,
\ee
where $m$ is the number of NS5-branes and
$\mathrm{LSM}_{4m}$ is the IR fixed point of a linear sigma model
which describes \kkl with  $K=4m$
and therefore central charge $6(1+~\!\!\frac{1}{m})$.

 Again, when we consider the case of 
$m=2$ NS5-branes, two transverse directions are described by a 
zero central charge $\frac{SU(2)}{U(1)}$ conformal theory, and
 thus play no role in the string theory dynamics. Therefore, the 
 $\mathcal{N}=1$ gauge theory for D4 and D6-branes attached to 
 these wrapped NS5-branes is equivalently described by D3 and
  D5-branes placed on $\mathbb{R}^{1,3} \times\,$\kkl.
  
Unfortunately, unlike the 2d cigar $\frac{SL(2,\mathbb{R})}{U(1)}$, 
the conformal theory describing \kkl is not known to be 
integrable~\cite{HoriKap}.
Nevertheless, 
this CFT has a mirror description in terms of the $\mathbb{Z}_m$ orbifold 
of a Landau-Ginzburg model with superpotential~\cite{HoriKap}:
\be\label{LG1}
W=e^{-mZ}\left(e^{-Y}+e^{Y}\right)+X^m\ ,
\ee
where $X$, $Y$ and $Z$ are chiral superfields and the $\mathbb{Z}_m$ action 
is generated by $Z\rightarrow Z-\frac{2\pi i}{m}$, $X\rightarrow 
e^{\frac{2\pi i}{m}}X$, while $Y$ is left invariant. 

From this point of view, it was argued in \cite{HoriKap} that
the conformal field theory description of this NS5-brane configuration 
allows for a straightforward deformation of \kkl to be constructed. 
The LG orbifold has a finite set of exactly marginal operators 
$e^{-(m-l)Z}X^l$, for $l=0,1,\ldots,m-2$. In particular for the 
case we are interested in, i.e. $m=2$, there is a one-parameter 
deformation of (\ref{LG1}):
\be\label{LG2}
W=e^{-2Z}\left(e^{-Y}+e^{Y}\right)+X^2+e^{t/2}\, e^{-2Z}\ .
\ee
From the point of view of the NS5-branes, this deformation corresponds to taking the two branes away one from the other while at the same time inflating the $\mathbb{CP}^1$ they wrap~\cite{HoriKap}. The deformed superpotential (\ref{LG2}) gives rise to a one-parameter deformation of the \kkl back\-ground. Remarkably all these spaces have the same large-radius behavior as \kkl (\ref{n2metric}). The two extrema of the family of deformations correspond to some special cases we already know: the $t\rightarrow -\infty$ limit obviously coincides with the unperturbed (\ref{LSM1}) configuration, while for $t\rightarrow +\infty$ we get the configuration corresponding to the superpotential \cite{HoriKap}:
\be
W=e^{-2Z'}+X^2\ ,
\ee
that is the mirror dual of two flat parallel NS5-branes.

Backed by these considerations, we argue that  each theory described  
by placing branes on the one-parameter family of vacua dual to  
(\ref{LG2}) could be  connected to the  $\mathcal{N}=2$ theory 
which is obtained by placing branes on $\mathbb{R}^{1,3}\times\mathbb{R}^2\times 
\frac{SL(2,\mathbb{R})_{4}}{U(1)}$.

D4-branes attached to the wrapped NS5-branes will have no possibility to 
move from their position when the NS5-branes wrap a vanishing cycle,
 while they will be free to move on a plane when the NS5-branes are flat. 
 In all other intermediate cases, their motion will be delimited to 
 the compact cycle wrapped by the NS5-branes. Since the position of the 
 D4-branes on the NS5-branes is usually described by  adjoint fields  
 of the dual field theory, these 
 will have a superpotential that reproduces the 
 above behavior. 

As we argued at the end of section \ref{subsect: stability}, our $AdS_5\times
\tilde{S}^3$ back-reacted solution is expected to have a massless mode in its spectrum, corresponding to a marginal operator in the dual field theory. In view of the above description, we can interpret this massless mode as indicating on which one of the one-parameter family of deformed \kkl vacua we have placed D3 and D5-branes.

We report here also  the following curious fact which, for the moment,
lacking a deeper understanding of branes in non-critical string
theory, we cannot consider as anything more than  an intriguing accident. 
Our 8d background has the form:
\be
      ds^2 = \frac{R_{AdS}^2}{z^2}(dx^2_{1,3}+dz^2) + R_{AdS}^2\,ds_3^2\ ,
\ee
where $ds_3^2$ is the metric of a 
supersymmetric three-cycle of $T^{1,1}$:
\be
ds_3^2 =
\frac{1}{9}(d\psi+\cos\theta d\varphi)^2+\frac{1}{6}(d\theta^2+\sin^2\theta 
d\varphi^2)\ .
\ee
We have to remark, anyway, that $\alpha'$ corrections could well change 
the values
of the eight-dimensional metric coefficients, and make the similarity 
disappear. If not, this could suggest a relation between the SCFT 
dual of our solution and the quiver Klebanov-Witten 
model dual to string theory on $AdS_5\times T^{1,1}$ \cite{KW}.

The problem of identifying the dual field theory would receive crucial
hints from the study of branes on \kkl in a way similar to what was done for the
cigar in \cite{FotoNiarPrez, Murthy}. As we have already noticed, the conformal
theory describing \kkl may not be integrable, and therefore this approach might
prove very difficult.  Nonetheless, considering the problem 
from the perspective of the dual 
Landau-Ginzburg description
 could be useful. 

Let us  conclude this section with some brief symmetry considerations.
As a first observation, note that there is only one kind of D3-wrapped-on-$\tilde{S}^3$ we can construct in our
case, i.e. only one kind of baryon vertex in the dual field theory. 
 This requires the dual field theory to have baryons all with the same
conformal dimension.
Concerning the non-Abelian flavor symmetry of the field theory, 
it is realized in the open string sector,
while the baryonic $U(1)_B$ should be dual, as usual, 
to the reduction of the four-form potential on $\tilde S^3$, 
the latter identification being in agreement with the baryon vertex just described.
Finally, we can always switch on a constant axion on our background, 
playing the role of the $\theta_{YM}$ angle.

\section{Fundamental flavors and problems for holography}\label{gen probls}

The non-critical backgrounds found in this paper should {\it not}
 be taken as  faithful {\it holographic gravity} duals of the corresponding ${\cal N}=0$ and ${\cal N}=1$ conformal field theories. 
 In this sense these solutions suffer from the same limitations as those 
  in \cite{KlebMald}. These are due to the intrinsic 
 limits of the two-derivative approximation to non-critical string theory, as for example the fact that the backgrounds 
 have generically large curvature, or the lack of proofs that some
   supersymmetry is really preserved. 

There is a simple known reason why we expect that conformal theories with fundamental (and antifundamental) flavors cannot have a weakly curved holographic dual. This is related to the structure of their gravitational central charges $a$ and $c$. We examine this structure in the supersymmetric case where we can make use of exact results. In  ${\cal N}=1$ terms, the gravitational central charges can be generically written as combinations of certain 't Hooft anomalies of the exact R-symmetry \cite{anselmi}:
\be
a = {3\over32}[3\mathrm{Tr}\,R^3-\mathrm{Tr}\,R]\ , \quad c={1\over32}[9\mathrm{Tr}\,R^3-5\mathrm{Tr}\,R]\ ,
\label{aec}
\ee
so that:
\begin{equation} 
a-c = \frac{1}{16} \mathrm{Tr}\, R\ .
\end{equation}
For an ${\cal N}=1$ SCFT with gauge group $SU(N_c)$, $N_f$ chiral fields in the fundamental, $N_f$~chiral fields in the anti-fundamental and $N_A$ adjoint fields we have, at leading order in the large $N_c$ limit (so that $N_c^2-1\approx N_c^2$):
\bea 
\mathrm{Tr}\, R &=& N_c^2 + N_c^2\sum_{a=1}^{N_A} (R_a-1)+ N_c\sum_{f=1}^{N_f}(R_f-1) + N_c\sum_{{\tilde f}=1}^{N_f}(R_{\tilde f}-1)\ .
\label{trr}
\eea
Here we have used the fact that the ${\cal N}=1$ vector multiplet has R-charge 1 (so the gauge field has $R=0$), and we have labeled with $R_a$ the R-charge of the adjoint multiplets and with $R_f$, $R_{\tilde f}$ the ones of the fundamentals and anti-fundamentals. 

Combining (\ref{trr}) with the relation imposed by the vanishing of the NSVZ beta function, it is easy to deduce the explicit expression:

\begin{equation}\label{amenc}
a-c = \frac{N_c}{32}[\sum_{f=1}^{N_f} (R_f-1)+\sum_{{\tilde f}=1}^{N_f}(R_{\tilde f}-1)]\ .
\end{equation}
This result is a non-perturbative, exact consequence of the exact formulae for the central charges and the $\beta$ function, at large $N_c$, so that, in particular, we can compare it with dual string theory results. Notice that the adjoint fields never contribute to this expression.

From the formula above it follows that in a generic SCFT with fundamentals and anti-fundamentals, the non-zero difference $a-c$ may be leading in the large $N_c$ limit. This is indeed what happens for the ${\cal N} = 1,2 $ conformal SQCD's \cite{anselmidos}: for ${\cal N}=2$ conformal SQCD, at leading order, $a={7\over24}N_c^2$ and $c={N_c^2\over3}$, where $N_f$ and $N_c$ are of the same order of magnitude; for ${\cal N}=1$ conformal SQCD $a={3\over8}N_c^2[1-{3\over2}{N_c^2\over N_f^2}]$, $c={7\over16}N_c^2[1-{9\over7}{N_c^2\over N_f^2}]$.

Here we see a crucial difference from what happens in standard critical $AdS$/CFT examples, where the SCFT's have matter fields only in the bifundamental, adjoint and (anti)symmetric representations. In those cases, $\mathrm{Tr}R$ is subleading in $N_c$ in the large $N_c$ limit. This fact exactly matches with the holographic prediction $a=c$ from the dual $AdS_5\times X_5$ weakly curved backgrounds \cite{HS}. 

This holographic prediction automatically extends to every theory which has a classical weakly curved dual $AdS_5\times X_q$ background with constant dilaton $\phi$:
\be
a=c \approx V(X_q)e^{-2\phi}R_{AdS}^3\ ,
\label{a-c}
\ee
where $V(X_q)$ is the volume of $X_q$ and $R_{AdS}$ is the $AdS$ radius. 

This result implies that superconformal theories with $N_f=O(N_c)$ 
fundamental fields
cannot have a purely supergravity weakly curved holographic 
dual\footnote{We are grateful to J. Sonnenschein, who  was  
independently studying the same problems, for several discussions 
on the matter.}. This is in agreement with what is found in this 
paper and in \cite{KlebMald}: in both cases the supergravity 
backgrounds have large curvatures. 

This argument can be
extended to weakly coupled exactly conformal non-supersymmetric theories,
for which the $\beta$-functions, $a$ and $c$ can be written in terms
of the number of each kind of fields present in the theory. Although
it is natural to expect that the same conclusion applies to general
non-supersymmetric conformal fixed points, an analogous proof cannot 
be given in such cases.

In order to compute the value of the central charges of the  theories above from the dual non-critical string models, one would therefore have to go beyond the supergravity limit.
Just as a curiosity, formula (\ref{a-c}) applied to the eight-dimensional background studied in this paper gives $a=c \approx N_c^2$.
This result could eventually capture the correct scaling, but not the 
actual coefficients, of the central charges of a SCFT where $N_f/N_c$ is finite.

Apart from the central charges of the theory, other protected SCFT quantities that could be hopefully deduced from the dual non-critical string background are the baryon charges. Their stringy evaluation does not seem to suffer from the same intrinsic limitations encountered for the holographic calculation of the central charges. Just to focus on the supersymmetric example of this paper, it is expected that the baryonic $U(1)_B$ symmetry of the dual ${\cal N}=1$ SCFT is holographically realized by the abelian field in $AdS$ coming from the reduction of $C_4$ on the squashed three-sphere. Then, field theory baryons should be dual to D3 branes wrapped on ${\tilde S}^3$. The baryons are gauge invariant antisymmetrized products of $N_c$ fundamental or anti-fundamental fields. Hence their conformal dimension is proportional to $N_c$. A wrapped D3 brane on ${\tilde S}^3$ has an effective mass given by:
\be
m=T_3 V({\tilde S}^3)e^{-\phi}= T_3 16\pi^2 Q_c\ ,
\ee 
where $T_3$ is the D3-brane tension. In the large mass approximation the conformal dimension of the dual operator is given by:
\be
\Delta=R_{AdS}m = R_{AdS} T_3 16\pi^2 Q_c\ .
\ee
This result could reproduce the correct scaling of $\Delta$ with $N_c$, 
and could be used in principle to guess the value of $R_{AdS}$ in the 
$\alpha'$ corrected solution provided the exact dual SCFT is identified. 

\section{Summary and outlook}\label{summary}

In the last few years it has been realized that faithful string duals of some phenomenologically interesting theories,
like pure Yang-Mills or QCD, involve string theories in less than ten dimensions.

In general, 
non-critical solutions have high curvature at the order of the string scale.
This makes the task really difficult, since one should consider the
full string theory in back-reacted,
highly curved backgrounds with RR fields. This seems far from our
present possibilities, but there have been some steps forward using
two complementary approaches: in the first one \cite{FotoNiarPrez,Murthy},
world-sheet CFT techniques are used to study the physics of branes in
the (non-backreacted) cigar, the relevant geometry to engineer the string dual of
${\cal N}=1$ SYM. In the second one
\cite{KlebMald,KupSon,KupSon2}, the two-derivative approximation is used
in the belief that stringy corrections should not spoil the qualitative
behavior of  gravity solutions which are highly symmetric (like
$AdS$ solutions). This is of course an old 
argument \cite{cave,Poly}, but only recently it has been 
applied to theories with a large number of fundamental fields. 

In this paper, we have further pursued this second approach. 
First, we have proposed a five-dimensional Anti deSitter dual of conformal QCD
by considering stacks of D3 color branes and D4-anti D4 flavor branes.
The limits of the conformal window should correspond to the values
of $\frac{N_f}{N_c}$ for which the dimension of the scalar meson
operators hit the unitarity bound. We have also provided an interpretation
of these meson operators from the world-volume dynamics of the flavor 
branes.

Afterwards, we have explored the possibility of constructing a 
holographic dual by considering branes on a previously known 
four-dimensional space which we have dubbed KKL$_4$.
Placing D3 branes at the tip of $\mathbb{R}^{1,3}\times$ KKL$_4$ along
with D5 flavor branes appropriately embedded in the geometry, we have
engineered the dual of an ${\cal N}=1$ 
conformal gauge theory with $N_f$ chiral
multiplets in the fundamental and anti-fundamental representations.
Our results suggest that there is a supersymmetric, conformal
($AdS_5 \times \tilde S^3$) solution for just one value of the ratio
$\frac{N_f}{N_c}$.  Finding the corresponding dual SCFT remains a challenge.
It would be very interesting to study branes in KKL$_4$ {\`a} la
\cite{FotoNiarPrez, Murthy}, since this would provide  crucial information on the dual gauge theory.

As a general remark, we have provided further evidence that the two
derivative approximation to non-critical strings can give some insight
and qualitative agreement in the study of holographic pairs.
However, it is not enough to get quantitative information about the 
dual field theories. For instance (see section \ref{gen probls}),
a difference of gravitational central charges $a-c$ must come in
the string dual from higher derivative terms \cite{HS}. Since 
$a-c \neq 0$ in  conformal theories with fundamental flavors, it is 
of obvious interest to develop tools that may allow us to study better such
string duals.

\acknowledgments

\noindent It is a pleasure to thank M. Bianchi,
A. Fotopoulos, D. Israel, L. Martucci, S. Murthy, V. Niarchos,
C. N\'u\~nez,
M. Petropoulos, N. Prezas, K. Skenderis, J. Sonnenschein and J. Troost 
for valuable discussions. A. L. C. would like to thank Ecole Polytechnique 
for support during the preparation of this paper and Pascal Anastasopoulos 
for his help with citations. This work was partially supported by
INTAS grant, 03-51-6346, RTN contracts MRTN-CT-2004-005104 and
MRTN-CT-2004-503369, CNRS PICS \#~2530 and 3059,
by MCYT FPA 2004-04582-C02-01, CIRIT GC 2001SGR-00065
 and by a European Union Excellence Grant,
MEXT-CT-2003-509661.

\appendix

\setcounter{equation}{0}
\renewcommand{\theequation}{\Alph{section}.\arabic{equation}}

\section{Explicit check of the 8d Einstein equations}
\label{Einsteinsec}

The aim of this appendix is to write explicitly the set of 
relevant Einstein equations in the 8d case. Let us switch, for convenience, to 
Einstein frame $g_{\mu\nu}=e^{\frac{2}{3}\phi}g_{\mu\nu}^E$. 
The action (\ref{d=8 action}) now reads:
\be\label{Einsaction}
S=\frac{1}{2 \kappa_{(8)}^2}\int d^8x \sqrt{-g^E_{(8)}}
\left(R-\frac{2}{3}(\partial_\mu \phi)^2 + 2 e^{\frac{2}{3}\phi} - 
e^{-\frac{2}{3}\phi} F_{(5)}^2 \right) + S^E_{flavor}\,\,,
\ee
where the action of the flavor branes is:
\be
S^E_{flavor}=-{T_5 N_f \over 4\pi}
 \int d^8x \sin\theta \,e^\phi \sqrt{-\hat g^E_{(6)}}\,\,.
\label{Einsflavact}
\ee
We denote by $\hat g^E_{(6)}$ the determinant of the metric induced on the
D5 flavor branes, which is simply:
\be
d\hat s_E^2=e^{-\frac{2}{3}\phi}
\left( e^{2\lambda}\,dx_{1,3}^2+d\tau^2
+e^{2f}d\psi^2\right)\,\,.
\label{inducedmetric}
\ee
The Einstein equations read:
\bear
&R_{\mu\nu}-\frac{1}{2} g_{\mu\nu} R = \frac{2}{3}(\partial_\mu \phi)
(\partial_\nu \phi)-\frac{1}{3} g_{\mu\nu}(\partial \phi)^2 + g_{\mu\nu} \,
e^{\frac{2}{3}\phi}+&\rc\rc
&+\,\frac{1}{5!}e^{-\frac{2}{3}\phi}
\left(5 F_{\mu\lambda_2 \lambda_3\lambda_4\lambda_5}
F_\nu^{\ \lambda_2 \lambda_3\lambda_4\lambda_5}-\frac{1}{2}
g_{\mu\nu} F_{\lambda_1\lambda_2 \lambda_3\lambda_4\lambda_5}
F^{\lambda_1\lambda_2 \lambda_3\lambda_4\lambda_5}\right)+
T_{\mu\nu}^{flavor}\,\,.&
\label{Einsteqs}
\eear
The stress-energy tensor of the D5 branes can be computed as:
\be
T^{\mu\nu}_{flavor}={2 \kappa_{(8)}^2 \over \sqrt{-g^E_{(8)}}}\,
{\delta {\cal L}^E_{flavor} \over \delta {(g_{\mu\nu}^{(8)})_E}}=
-{\kappa_{(8)}^2 T_5 \over 2\pi} N_f \sin\theta\,
e^\phi\, \frac{1}{2}\,
\delta_i^\mu \delta_j^\nu \,(\hat g_E^{(6)})^{ij}\sqrt{
\hat g^E_{(6)} \over g^E_{(8)}}\,\,,
\label{Tmunuup}
\ee
where latin indices span the six dimensional world-volume.
Inserting (\ref{Qfdefinition}), one can readily write the 
components of the flavor stress-energy tensor:
\bear
&T_{x_a x_b}= -{Q_f \over 2} e^{2\la +\phi -2\nu}\eta_{ab}\spa\quad
&T_{\tau\tau}= -{Q_f \over 2} e^{\phi -2\nu}\spa\rc
&T_{\psi\psi}= -{Q_f \over 2} e^{\phi +2f -2\nu}\spa\qquad\quad
&T_{\varphi\psi}=-{Q_f \over 2} e^{\phi+2f -2\nu}\cos\theta\spa\rc
&T_{\varphi\varphi}=-{Q_f \over 2} e^{\phi+2f -2\nu}\cos^2 \theta\,\,.&
\label{Tmunu}
\eear
One can also obtain the equation of motion of the dilaton from
the action (\ref{Einsaction}):
\be
{1\over {\sqrt{-g^E_{(8)}}}}\partial_\mu \left(g_E^{\mu\nu} \sqrt{-g^E_{(8)}}
\partial_\nu \phi\right)+ e^{\frac{2}{3}\phi}+
\frac{1}{2}e^{-\frac{2}{3}\phi}F_{(5)}^2-\frac{3}{4}
Q_f e^{{5\phi\over 3}
-2\nu}\,=\,0\,\,.
\label{dilatoneq}
\ee
Finally, the equation for the five-form is:
\be
\partial_\mu \left(\sqrt{-g_{(8)}^E} e^{-\frac{2}{3}\phi}
F^{\mu\lambda_2 \lambda_3\lambda_4\lambda_5}\right)\,=\,0\,\,.
\label{formeq}
\ee
It is straightforward to verify that equations
(\ref{Einsteqs}), (\ref{dilatoneq}), (\ref{formeq})
 are satisfied by the general solution
displayed in section \ref{secgeneral} as well as by  the
$\RR^{1,3} \times$ \kkl vacuum
(\ref{n2metric}), once the respective
metrics are transformed to the Einstein frame.

It is worth making a technical comment on the construction of the
solution.
Notice that the squashed three-sphere is not an Einstein space.
In fact, the
angular dependence in its 
Ricci tensor  is not proportional to that in  
the metric tensor $g_{\mu\nu}$.
However, for the given ansatz, all other terms in 
(\ref{Einsteqs}) except for the flavor one are indeed proportional
to $g_{\mu\nu}$. Therefore, in order to satisfy the Einstein equations,
$T_{\mu\nu}^{flavor}$ cannot have the same angular dependence as
$g_{\mu\nu}$, as it is of course the case in (\ref{Tmunu}).
 This is an effect of the smearing
and would not happen for space-time filling D7  branes.
Therefore, we conclude that an $AdS_5 \times 
\tilde S^3$ solution cannot be obtained 
by just using D7 instead of D5.

Finally, let us notice that when four-dimensional gauge theories are engineered
in the ten dimensional IIA theory with branes suspended between
branes, flavors are usually related to D6-branes. Since the non-critical
backgrounds are related to the double scaling limits of such theories,
it is much more natural to have in our setup
D5-flavor branes rather than D7.

\section{Derivation of the $\mathbb{R}^2\times\,$cigar and \kkl solutions from first order BPS equations}\label{app: KKL from super}

In this appendix we obtain the \kkl vacuum \cite{KKL} as a solution of the first order BPS equations we derive from the vacuum superpotential (\ref{vacuum super}), i.e. we take $Q_c=Q_f=0$:
\be\label{vacuum super app}
W=-\frac{1}{2}e^{\frac{2}{3}\phi-2\gamma}-e^{-2\beta+2\gamma}-\frac{1}{2}e^{-2\beta-4\gamma}\ .
\ee
We introduce the radial coordinate $\zeta$ such that $d\zeta=4\,e^{\frac{1}{3}\phi+\beta-2\gamma}d\tau$, 
and use the same parametrization of the metric as in (\ref{8dmetric}); the relation of this parametrization  to the fields in (\ref{vacuum super app}) is shown in (\ref{repar flow}). The BPS first order equations  read then:
\be\label {vacuum BPS}
\begin{split}
&\partial_\zeta\phi=-\frac{1}{8}\ ,\\
&\partial_\zeta\nu=\frac{1}{8}e^{-2\nu}\ ,\\
&\partial_\zeta f=-\frac{1}{8}+\frac{1}{4}e^{-2f}-\frac{1}{8}e^{-2\nu}\ ,\\
&\partial_\zeta \lambda=0\ .
\end{split}
\ee
The equations for the fields $\phi$, $\nu$ and $\lambda$ are trivial; 
remarkably, also the equation for $f$ admits an analytic solution. We find:
\be\label{sol BPS vac}
\begin{split}
& \phi(\zeta)=\phi_0-\frac{1}{8}\zeta\ ,\\[3pt]
&e^{2\nu(\zeta)}=e^{2\nu_0}+\frac{\zeta}{4}\ ,\\[3pt]
&e^{2f(\zeta)}=8\frac{\frac{C}{8}e^{-\frac{\zeta}{4}}+\frac{\zeta}{4}-1+e^{2\nu_0}}{\zeta+4e^{2\nu_0}}\ ,\\[3pt]
& \lambda(\zeta)=\lambda_0\ ,
\end{split}
\ee
which is well defined only for:
\be
\zeta\geq -4e^{2\nu_0}\ .
\ee
The integration constant $\nu_0$ fixes the origin of the radial coordinate $\zeta$, so we can fix it in such a way that $e^{2\nu_0}=0$ and $\zeta\geq 0$. $\phi_0$ is the value of the linear dilaton at $\zeta=0$ and $\lambda_0$ is a scale factor which we can fix to $1$ by a rescaling of the $\mathbb{R}^{1,3}$ coordinates. Finally the constant $C$ has to be fixed to $8$ in order for the solution to be regular at $\zeta=0$.

The only thing that remains to be done is  to evaluate  the coefficient in front of the term $d\zeta^2$ in the metric. From the definition of $\zeta$, $d\zeta=4\,e^{\frac{1}{3}\phi+\beta-2\gamma}d\tau=4\, e^f d\tau$, and the solution (\ref{sol BPS vac}), it is easy to show that $d\tau^2=\frac{\zeta}{128(e^\frac{-\zeta}{4}+\frac{\zeta}{4}-1)}d\zeta^2$. Collecting everything together, we can write the general solution to the vacuum BPS equations (\ref{vacuum BPS}):
\be
\begin{split}
ds^2=\;&dx_{1,3}^2+\frac{1}{4}\,\frac{\zeta}{32(e^\frac{-\zeta}{4}+\frac{\zeta}{4}-1)}d\zeta^2+\frac{\zeta}{4}(d\theta^2+\sin^2\theta d\varphi^2)+\\&+\frac{1}{4}\,\frac{32(e^\frac{-\zeta}{4}+\frac{\zeta}{4}-1)}{\zeta}(d\psi+\cos\theta d\varphi)^2\ ,\\[3pt]
\phi(\zeta)=&\phi_0-\frac{\zeta}{8}\ .
\end{split}
\ee
This is exactly (\ref{n2metric}) with $K=8$, which is the right value to cancel the central charge  for a background of the form $\mathbb{R}^{1,3}\times$ \kkl (see (\ref{cc})).

Going back to  (\ref{vacuum BPS}), we can also consider a 
somehow singular  solution of the first order equations.
One can take $\nu(\zeta)=+\infty$ which corresponds 
to looking for a solution for which the $S^2$ base of the $\tilde{S}^3$ 
fibration has been blown up to a flat plane and therefore the squashed 
three-sphere has been substituted by the direct product of a plane 
and a circle parameterized by the angle $\psi$, with radius $e^{f}$. 
The equations for $\phi(\zeta)$ and $\lambda(\zeta)$ are not altered, 
so their solutions will be the same as in (\ref{sol BPS vac}). The equation for $f(\zeta)$, instead, now reads:
\be
\partial_\tau e^{-f}=\frac{1}{2}-e^{-2f}\ ,
\ee
where we have gone back to the radial coordinate $\tau$. The solution of this equation is:
\be
e^{f(\tau)}=\sqrt{2}\,\tanh \frac{\tau}{\sqrt{2}}\ .
\ee
Rewriting the whole solution in terms of the variable $\tau$ we then have:
\be
\begin{split}
&ds^2=dx^2_{1,3}+dy_1^2+dy_2^2+d\tau^2+2\tanh^2\frac{\tau}{\sqrt{2}} \;d\psi^2\ ,\\
&e^{2\phi(\tau)}=\frac{e^{2\phi_0}}{\cosh^2 \frac{\tau}{\sqrt{2}}}\ ,
\end{split}
\ee
which is the $\mathbb{R}^{1,3}\times \mathbb{R}^2\times \frac{SL(2,\mathbb{R})}{U(1)}$ we claim to be the eight-dimensional vacuum where one needs to put branes to obtain the non-critical string dual of $\mathcal{N}=2$ QCD. Notice that the BPS equations give us the cigar with the right CFT  level to be a good eight-dimensional background of non-critical string theory.

\section{Existence of a superpotential for  generic $N_f/N_c$}\label{app: exist W}

We demonstrate in this appendix that a superpotential which solves equation (\ref{diffW}) should exist for all values of $N_f/N_c$. This is an adaptation of an analogous argument of \cite{KupSon}, where it is applied to backgrounds of the form $AdS_n\times S^k$, with $S^k$ a round $k$-sphere.

Let us start from the effective action written as in (\ref{Seffu}); the superpotential $W$ has then to satisfy equation (\ref{diffW}):
\be\label{diffWapp}
\frac{3}{8}\left(\frac{\partial W}{\partial \phi}\right)^2 +\frac{1}{24}\left(\frac{\partial W}{\partial \beta}\right)^2 +\frac{1}{24}\left(\frac{\partial W}{\partial \gamma}\right)^2-\frac{1}{3}W^2=V(\phi,\beta,\gamma)\ .
\ee
If a solution to this partial differential equation exists, that is if a superpotential $W$ can be defined, then our general $AdS_5\times\tilde{S}^3$ solution (\ref{solut beta}) must also solve the first order BPS equations of motion associated to $W$. In particular, then, we see from (\ref{BPSgeneral}) that this implies:
\be
\left.\frac{\partial W}{\partial \phi}\right|_{0}=0\ ,\qquad\quad\left.\frac{\partial W}{\partial \beta}\right|_{0}=0\ ,\qquad\quad\left.\frac{\partial W}{\partial \gamma}\right|_{0}=0\ .
\ee
where $|_0$ stands for the value of the function at one of the $AdS_5\times\tilde{S}^3$ solutions.

We notice that the choice of fields we have made is particularly convenient: the potential and the superpotential only depend on those fields which are set to a constant for solutions of the form $AdS_5\times\tilde{S}^3$. By expanding both sides of (\ref{diffWapp}) around the solution (\ref{solut beta}) we obtain the following equation for the derivatives of the superpotential at the $AdS$ points:
\be
\begin{split}
\left.V\right|_{0}+\ldots=&\frac{3}{8}\left.\frac{\partial W}{\partial \phi}\right|^2_{0}+\frac{1}{24}\left.\frac{\partial W}{\partial \beta}\right|^2_{0}+\frac{1}{24}\left.\frac{\partial W}{\partial \gamma}\right|_{0}^2-\frac{1}{3}\left.W\right|^2_{0}+2(\phi-\phi_0)\left(\frac{3}{8}\left.\frac{\partial W}{\partial \phi}\right|_{0}\left.\frac{\partial^2 W}{\partial \phi^2}\right|_{0}+\right.\\[5pt]
&+\left.\frac{1}{24}\left.\frac{\partial W}{\partial \beta}\right|_{0}\left.\frac{\partial^2 W}{\partial \phi\partial\beta}\right|_{0}+\frac{1}{24}\left.\frac{\partial W}{\partial \gamma}\right|_{0}\left.\frac{\partial^2 W}{\partial \phi\partial\gamma}\right|_{0}-\frac{1}{3}\left.W\right|_{0}\left.\frac{\partial W}{\partial\phi}\right|_{0}\right)+\\[5pt]
&+2(\beta-\beta_0)\left(\frac{3}{8}\left.\frac{\partial W}{\partial \phi}\right|_{0}\left.\frac{\partial^2 W}{\partial \beta\partial\phi}\right|_{0}+\frac{1}{24}\left.\frac{\partial W}{\partial \beta}\right|_{0}\left.\frac{\partial^2 W}{\partial\beta^2}\right|_{0}+\frac{1}{24}\left.\frac{\partial W}{\partial \gamma}\right|_{0}\left.\frac{\partial^2 W}{\partial \beta\partial\gamma}\right|_{0}+\right.\\[5pt]
&-\left.\frac{1}{3}\left.W\right|_{0}\left.\frac{\partial W}{\partial\beta}\right|_{0}\right)+
2(\gamma-\gamma_0)\left(\frac{3}{8}\left.\frac{\partial W}{\partial \phi}\right|_{0}\left.\frac{\partial^2 W}{\partial \gamma\partial\phi}\right|_{0}+\frac{1}{24}\left.\frac{\partial W}{\partial \beta}\right|_{0}\left.\frac{\partial^2 W}{\partial\gamma\partial\beta}\right|_{0}+\right.\\[5pt]
&+\left.\frac{1}{24}\left.\frac{\partial W}{\partial \gamma}\right|_{0}\left.\frac{\partial^2 W}{\partial \gamma^2}\right|_{0}-\frac{1}{3}\left.W\right|_{0}\left.\frac{\partial W}{\partial\gamma}\right|_{0}\right)+\ldots\ ,
\end{split}
\ee
where $\ldots$ stand for quadratic terms, which are irrelevant for the present analysis. By matching the terms on both sides of the above equation, it is easy to show that a solution is given by:
\be\label{solut xpans}
\left.\frac{\partial W}{\partial \phi}\right|_{0}=\left.\frac{\partial W}{\partial \beta}\right|_{0}=\left.\frac{\partial W}{\partial \gamma}\right|_{0}=0\ ,\qquad\quad-\frac{1}{3}\left.W\right|_0^2=\left.V\right|_0\ .
\ee
Notice that the vanishing of the first derivatives of the superpotential at the $AdS_5\times\tilde{S}^3$ solution does not depend on the specific value of the parameter $A$ determining the solution, that is this argument is valid for every value of the ratio $N_f/N_c$. As a final check, let us evaluate the $AdS$ radius. The BPS equation that fixes the radius of $AdS$ is:
\be
\partial_u \alpha=\partial_u\left(-\frac{2}{3}\phi+\frac{2}{3}\nu+\lambda+\frac{1}{3}f\right)=-\frac{1}{3}W\ .
\ee
Since all fields are constant in the $AdS$ solution but $\lambda(u)$, the above equation gives us the value of $\partial_u\lambda$. We only need to calculate the constant value of W as given by the last equation in (\ref{solut xpans}) and switch back to the  radial coordinate $\tau$ to compare our result with (\ref{AdS repar}). We find:
\be
k_0=\partial_\tau\lambda=e^{-\frac{2}{3}\phi+\frac{2}{3}\nu+\frac{1}{3}f}\partial_u\lambda=\frac{1}{2\sqrt{3}}\sqrt{\frac{A+3}{A+1}}\ ,
\ee
which agrees with the value in (\ref{AdS repar}) we had obtained through the second order equations of motion.

\section{Semiclassical analysis of D5-branes in $\RR^{1,3}\times $ \kkl}\label{appendi}

Some D-branes embedded in \kkl space were studied in \cite{Nakayama}.
Unfortunately, they do not include the kind of D5-branes that we are
using to introduce flavor. 

In this appendix we make a semiclassical analysis, based on the
Dirac-Born-Infeld action of these branes. This is similar to what was 
done for D2 branes on the cigar \cite{Fotop}, which was shown to be 
in good qualitative agreement with the exact solution \cite{RibSchom,Niarchos}.

Let us consider a D5-brane extended along $x_0,\dots,x_3,\zeta,\psi$
inside the geometry (\ref{8dvacuum}), (\ref{n2metric}) at constant
$\theta$, $\varphi$. It is easy to prove that this embedding is consistent
with the DBI action. We then have, considering the excitation of
the world-volume gauge field on the directions inside \kkl:
\be
{\cal L}=  -T_5 e^{-\phi}\,\sqrt{-\det(P[g]+2\pi F)}=
-T_5\, e^\frac{\zeta}{8} \sqrt{\frac{1}{16}+(2\pi F_{\zeta\psi})^2}\,\,.
\ee
The equation of motion for $A_\psi$ is ${\partial {\cal L}
\over \partial F_{\zeta\psi}}=const$ which renders:
\be
2\pi F_{\zeta\psi}={1 \over 4\sqrt{e^{\frac{\zeta-\zeta_0}{4}}-1}}\,\,,
\label{Ffield}
\ee
where $\zeta_0$ is a constant of integration. There are two, qualitatively
different cases, for $\zeta_0 < 0$ and $\zeta_0 \geq 0$. In
the first case, the D5 is extended up to the tip of the space
$\zeta = 0$. In the second one, analogously to the D2-branes on
the cigar considered in \cite{Niarchos}, the D5 terminates at a 
finite distance $\zeta_0$ from the tip. This can be seen by noticing
the vanishing of $T_{\zeta\zeta}=F_{\zeta\psi}{\partial {\cal L}
\over {\partial F_{\zeta\psi}}}-{\cal L}$ at that point.
A non-vanishing $F$-field implies that the D5 brane carries some effective
D3-brane charge, the total amount of which is given by 
$\int F_{\zeta\psi} d\zeta d\psi$. One can think of this object
as a D3-D5 bound state.
Since $F$ vanishes for $\zeta \to \infty$,
this D3 charge is concentrated at the bottom of the D5. This is
also similar to the cigar case \cite{RibSchom}.

Notice that, for any value $\zeta_0 \geq 0$, the effective D3 charge is
always the same. Furthermore, it is easy to prove that the energy of such
configuration $E=\int - {\cal L}$, although infinite, does not depend
on $\zeta_0$, which is therefore a marginal deformation in this case
(the same is not true for $\zeta_0 < 0$).

\end{document}